\documentclass[journal,twoside,web]{ieeecolor2}

\let\labelindent\relax

\usepackage{generic}
\usepackage{cite}
\usepackage{amsmath,amssymb,amsfonts}
\usepackage{algorithmic}
\usepackage{graphicx}
\usepackage{enumitem}
\usepackage{textcomp}
\usepackage{subcaption}

\usepackage{float} 
\usepackage{booktabs}
\usepackage{tikz}
\usepackage{xcolor}
\usepackage{amssymb}
\usepackage{hyperref}

\usepackage{multirow,cite,bm}
\usepackage{adjustbox}
\usepackage{url}
\usepackage{pifont}

\def\BibTeX{{\rm B\kern-.05em{\sc i\kern-.025em b}\kern-.08em
    T\kern-.1667em\lower.7ex\hbox{E}\kern-.125emX}}
\begin{document}
\title{TILES-2018 Sleep Benchmark Dataset: A Longitudinal Wearable Sleep Data Set of Hospital Workers for Modeling and Understanding Sleep Behaviors}

\author{
Tiantian Feng, \IEEEmembership{Member, IEEE}, Brandon M Booth, Karel Mundnich, Emily Zhou, \\ Benjamin Girault, Kristina Lerman and Shrikanth Narayanan \IEEEmembership{Fellow, IEEE}
\thanks{The research is based upon work supported by the Office of the Director of National Intelligence (ODNI), Intelligence Advanced Research Projects Activity (IARPA), via IARPA Contract No $2017$ - $17042800005$.}
\thanks{Tiantian Feng, Emily Zhou, and Shrikanth Narayanan are with Viterbi School of Engineering, University of Southern California, Los Angeles, CA 90089, USA (corresponding e-mail: tiantiaf@usc.edu; shri@usc.edu).}
\thanks{Kristina Lerman is with Information Sciences Institute (USC), Marina del Rey, CA, USA (e-mail:lerman@isi.edu).}
\thanks{Brandon M Booth is with the University of Memphis, Memphis, TN, 38152, USA (e-mail: brandon.m.booth@gmail.com); Karel Mundnich is with AWS AI Labs, USA (e-mail: kmundnic@gmail.com); Benjamin Girault is with Inria, France (e-mail: benjamin.girault@inria.fr). These authors performed the work while at USC.}
}

\maketitle

\begin{abstract}
Sleep is important for everyday functioning, overall well-being, and quality of life. 
Recent advances in wearable sensing technology have enabled continuous, noninvasive, and cost-effective monitoring of sleep patterns in real-world natural living settings. Wrist-worn devices, in particular, are capable of tracking sleep patterns using accelerometers and heart rate sensors.
To support sleep research in naturalistic environments using wearable sensors, we introduce the \textit{TILES-2018 Sleep Benchmark} dataset, which we make publicly available to the research community. This dataset was collected over a 10-week period from 139 hospital employees and includes over 6,000 unique sleep recordings, alongside self-reported survey data from each participant which include sleep quality, stress, and anxiety among other measurements.
We present in-depth analyses of sleep patterns by combining the \textit{TILES-2018 Sleep Benchmark} dataset with a previously released dataset, (TILES-2018), that follows a similar study protocol. Our analyses include sleep duration, sleep stages, and sleep diaries. Moreover, we report machine learning benchmarks using this dataset as a testbed for tasks including sleep stage classification, prediction of self-reported sleep quality, and classifying demographics. Overall, this dataset provides a valuable resource for advancing foundational studies in sleep behavior modeling.
\end{abstract}


\section{Introduction}
\label{sec:intro}

Healthy sleep is essential for maintaining physical and psychological well-being and cognitive functioning. Frequent sleep disturbances or insufficient sleep are often associated with sleep disorders, including insomnia  \cite{riemann2020sleep}, narcolepsy \cite{kornum2017narcolepsy}, and sleep apnea \cite{dempsey2010pathophysiology}. These sleep deficiencies may increase the risk of developing mental health disorders and chronic diseases. A recent article \cite{altevogt2006sleep} reports that approximately 50 to 70 million Americans experience sleep disorders of varying severity, reducing their well-being and productivity. The widespread recognition of the importance of sleep and the increasing prevalence of sleep disorders have spurred significant interest in understanding and improving sleep hygiene within the research community. 

\begin{table*}
    \footnotesize
    \centering

    \vspace{1mm}
    \scalebox{0.96}{
    \begin{tabular}{lcccccc}
        \toprule
        \textbf{Dataset} & 
        \multicolumn{1}{c}{\textbf{No. participants}} & 
        \multicolumn{1}{c}{\textbf{Study Duration}} & 
        \multicolumn{1}{c}{\textbf{Use of Wearable}} & 
        \multicolumn{1}{c}{\textbf{Device}} & 
        \textbf{Sleep Symptoms Measure} \\
        \midrule

        Sleep-EDF \cite{kemp2000analysis} & 100 & 2 nights & No & NA & Mild difficulty falling asleep \\
        2018 PhysioNet/CinC Challenge \cite{ghassemi2018you} & 1,985 & 1 night & No & NA & Unknown \\

        \midrule

        MESA \cite{chen2015racial} & 2,040 & 1 week & Yes & Actigraph & Sleep disordered breathing, etc.  \\
        ECSMP \cite{gao2021ecsmp} & 89 & 1 night & Yes & Empatica E4 & PSQI-based measures  \\
        DREAMT \cite{wang2024addressing} & 100 & 1 night & Yes & Empatica E4 & Snoring, Sleep apnea, etc.  \\

        \midrule
        
        TILES-2018 \cite{mundnich2020tiles} & 210 & 10 weeks & Yes & Fitbit Charge2 & PSQI-based measures  \\

        TILES-2018 Sleep Benchmark (this work) & 139 & 10 weeks & Yes & Fitbit Charge2 & PSQI-based measures  \\

        Combined TILES Sleep Dataset & 349 & 10 weeks & Yes & Fitbit Charge2 & PSQI-based measures  \\

        \bottomrule
    \end{tabular}
    }
    \caption{Several open-source datasets include sleep recordings. Our \textit{TILES-2018 Sleep Benchmark} dataset is one of the largest longitudinal wearable sleep datasets to date. It serves as a valuable resource for understanding sleep behaviors and evaluating computational models on real-world sleep data.}
    \vspace{-3.5mm}
    \label{tab:related_work}
\end{table*}

Conventional approaches for assessing sleep behaviors, such as polysomnography (PSG) \cite{rundo2019polysomnography, ibanez2018survey}, are typically conducted in clinical or laboratory settings. While PSG is considered the gold standard in sleep studies, it is often expensive, intrusive, and impractical for long-term monitoring over several weeks. In recent years, wearable sensors have emerged as a promising complement or alternative for measuring sleep behaviors in naturalistic settings \cite{sathyanarayana2016sleep, wang2018tracking, boe2019automating, ito2023effect, wang2024addressing}. These devices, typically worn on the wrist, are equipped with miniature sensors capable of recording physiological signals, such as heart rate. These wearable sensors provide continuous, non-invasive, minimally obtrusive, and cost-effective monitoring of sleep patterns, presenting unique opportunities to support individuals with sleep disorders.

In this paper, we introduce the \textit{TILES-2018 Sleep Benchmark} dataset\footnote{\href{https://tiles-data.isi.edu/}{https://tiles-data.isi.edu/}}, a longitudinal dataset collected using wearable sensors to explore sleep patterns in naturalistic, daily living settings, and benchmark frontier (pretrained deep learning) models for understanding sleep. Unlike previous sleep datasets that focus on laboratory settings, our datasets feature the collection of sleep data in natural settings over multiple weeks. Specifically, our sleep data are derived from the Fitbit Charge 2\footnote{https://www.fitbit.com/}, a commercial wearable sensor. We focus on studying the sleep patterns of hospital workers, who regularly face demanding work tasks and often work long shifts, including consecutive 12-hour and irregular shift schedules, such as night shifts \cite{ito2023effect, feng2021multimodal}. While ensuring continuous patient care, these work schedules introduce significant health risks, including various sleep disorders. For example, night shift nurses are particularly prone to shift work sleep disorder (SWSD) \cite{wright2013shift}, a circadian rhythm sleep disorder, due to the misalignment of work shifts with consistent sleep periods, such as nighttime and early morning.

The \textit{TILES-2018 Sleep Benchmark} dataset contains sleep statistics and continuous heart rate readings during sleep, collected from $139$ hospital employees over ten weeks. Overall, this dataset presents large-scale wearable sleep data, including over 6,000 unique sleep recordings and self-reported sleep quality scores (Table~\ref{tab:related_work}). When combined with the sleep data from our previously published \textit{TILES-2018} study, we are able to present several key analyses to understand sleep patterns using this data from 349 participants with approximately 15,000 unique sleep recordings. In addition, we establish the \textit{TILES-2018 Sleep Benchmark} dataset as the test partition to evaluate several machine learning tasks, including sleep-stage modeling, and prediction of self-reported sleep scores and demographics. 

Our contributions are summarized below:

\begin{itemize}
    \item We have collected, curated, and are releasing the \textit{TILES-2018 Sleep Benchmark} dataset\footnote{\href{https://tiles-data.isi.edu/}{https://tiles-data.isi.edu/}}, comprising wearable sleep recordings from $139$ hospital employees collected continuously over 10 weeks. The dataset also includes detailed demographics, self-reported PSQI scores, and a total of $6,012$ unique sleep recordings with continuous heart rate and device-provided sleep stages. The data access follows the same protocol described in the previous TILES-2018~\cite{mundnich2020tiles}.
    \item We report a comprehensive analysis that combines the current data with sleep data from our previous \textit{TILES-2018} dataset to examine distinct sleep patterns among hospital staff, focusing on the differences between primary work shift and primary work location.
    \item We design a set of machine learning experiments using the \textit{TILES-2018 Sleep Benchmark} dataset as the evaluation benchmark. These experiments include sleep-stage modeling, demographics classification, and prediction of self-reported PSQI scores. Our experiments demonstrate the unique value of the dataset to the machine learning and sleep informatics communities.
    
\end{itemize}

\begin{figure}[ht] {
    \centering
    \caption{The recruitment and study kick-off protocols of the \textit{TILES-2018 Sleep Benchmark} dataset.}
    
    \begin{tikzpicture}

        \node[draw=none,fill=none] at (0, 0){\includegraphics[width=0.95\linewidth]{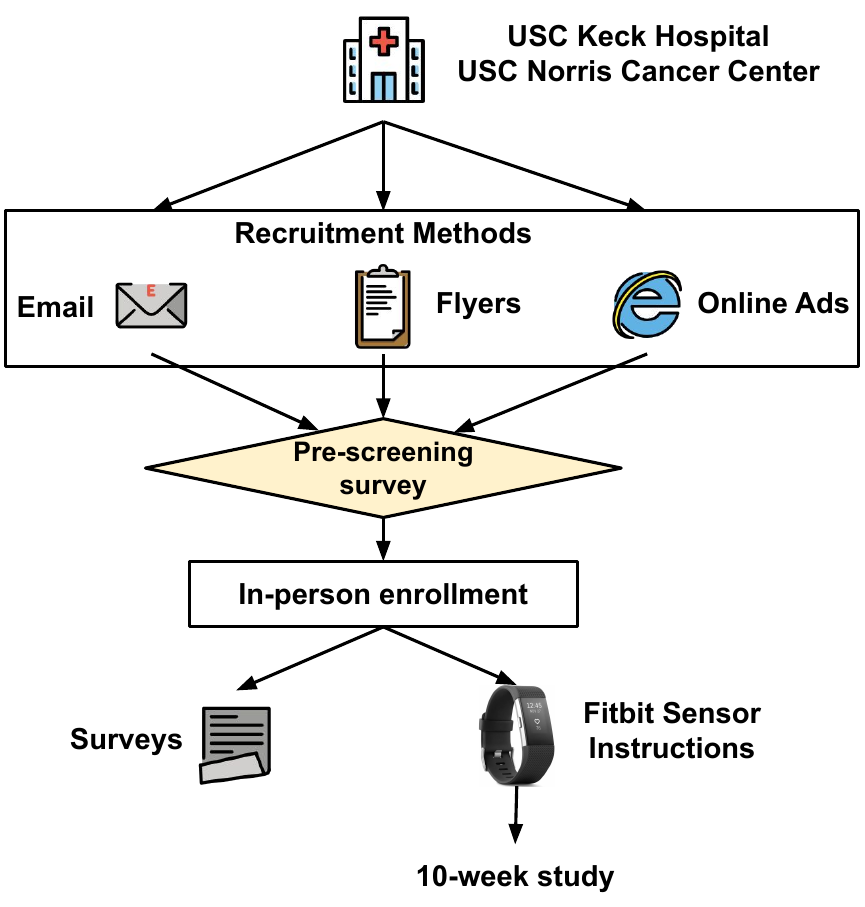}};

    \end{tikzpicture}
    
    \label{fig:study_protocol}
    \vspace{-3.5mm}
    
} \end{figure}

\section{Related Datasets}
\label{sec:related_work}

There have been numerous sleep-centered datasets focused on laboratory studies. For example, the Sleep-EDF dataset \cite{kemp2000analysis} is one of the most widely used data resource for sleep tracking that includes raw PSG signals to track detailed sleep stages and conditions. It contains EEG, ECG, and EOG signals from 197 unique recordings. Similarly, the 2018 PhysioNet/CinC Challenge dataset \cite{ghassemi2018you} includes 1,983 PSG recordings from patients evaluated for sleep disorders at the Massachusetts General Hospital. This dataset was used to develop algorithms for the automatic detection of sleep arousals from PSG data. While these datasets have made significant contributions and advances in understanding sleep, they are limited in capturing naturalistic sleep behaviors and offering insights into everyday sleep patterns.

\begin{table*}
    \centering
    
    \begin{tabular}{ccccc}

        \toprule
        
        \textbf{Folders} & \textbf{Subfolders} & \textbf{Subsubfolders} & \textbf{Description} & \textbf{File Split} \\
        \midrule

        \multirow{3}{*}{\texttt{fitbit}} & 
        \texttt{sleep-heart-rate} & & Heart rate time series (beats per minute) & per subject \\
        
        & \texttt{sleep-data} & NA & Sleep stages time series (classic labels or sleep stage labels) & per subject \\

        & \texttt{sleep-metadata} & & Sleep periods metadata & per subject \\
        \midrule

        \multirow{2}{*}{\texttt{fitbit}} & 
        \multirow{2}{*}{\texttt{participant-info}} & NA & IDs, demographics & \multirow{2}{*}{single file} \\

        & & & (age, sex, primary work shift, work position, primary work location) & \\
        \midrule
        
        \multirow{2}{*}{\texttt{surveys}} & \texttt{raw} & \texttt{baseline} & PSQI itemized answers & per subject \\
        & \texttt{scored} & \texttt{baseline} & PSQI scored answers & per subject \\
        
        \bottomrule

    \end{tabular}
    \caption{\textit{TILES-2018 Sleep Benchmark Data} Record. Three main folders contain wearable sleep data, self-reported sleep surveys, and participant metadata. The details of each data stream are included in each of the subfolders as README files.}
    \label{tab:dataset_record}
    \vspace{-1.5mm}
\end{table*}

In recent years, there has been a growing interest in collecting sleep data using wearable sensors, such as wristband-based devices. The Multi-Ethnic Study of Atherosclerosis (MESA) \cite{chen2015racial}, a longitudinal study examining subclinical cardiovascular disease across six U.S. communities, includes both one night of PSG recordings and a week of Actigraphy-based sleep data. Another example is the ECSMP dataset \cite{gao2021ecsmp}, which captures sleep data using Empatica E4 sensors but primarily recruited healthy college-based participants. More recently, the 2024 DREAMT dataset \cite{wang2024addressing} provides parallel recordings from both PSG and wearable Empatica E4 sensors to develop robust sleep modeling algorithms that generalize to different demographics. Despite these efforts to record sleep data using wearable devices, only a few studies have explored the collection of longitudinal sleep data across extended periods, leaving an important gap in understanding longer-term, real-world sleep behaviors. 


To this end, we introduce the \textit{TILES-2018 Sleep Benchmark} dataset, which provides 10 weeks of wearable sleep data collected from individuals working in hospital environments. Specifically, our datasets feature data collection in high-intensity work environments, such as ICU nursing units and medical labs, with varying shift schedules that include both day and night shifts, as well as outside work settings.

\section{Study Background and Logistics}
\label{sec:study}

\subsection{Study Background}

The \textit{TILES-2018 Sleep Benchmark} dataset is a part of the ``TILES: Tracking Individual Performance with Sensors`` project. This data was collected as an extended part of the \textit{TILES-2018} dataset collection effort and it was used as holdout data for evaluating the models and algorithms developed from the \textit{TILES-2018} dataset described in \cite{mundnich2020tiles}. Initiated in late 2018, this data collection involved acquiring a comprehensive set of physiological, environmental, and behavioral variables that may impact employee wellness and/or job performance. The present study reports original (hitherto unpublished) data that were collected over 10 weeks using wearable sensors, environmental sensors, and surveys from more than $139$ hospital employee volunteers who worked at the USC Keck Hospital and the USC Norris Comprehensive Cancer Center. Each participant provided written consent to participate in the study. All study procedures were conducted following the guidelines and approval of the University of Southern California's Institutional Review Board (study ID: HS-17-00876). The combined sleep data from this evaluation set and the data released in the previous TILES-2018 dataset \cite{mundnich2020tiles} provide a unique source of wearable sleep data collected in the wild that can be used to study sleep patterns among healthcare providers. We refer to the combination of the TILES-2018 Sleep Benchmark dataset and the sleep data in the TILES-2018 dataset as the \textit{Combined TILES Sleep} dataset.

\subsection{Study Logistics}
The study of the \textit{TILES-2018 Sleep Benchmark} dataset adheres to the logistics outlined in our previously released \textit{TILES-2018} dataset \cite{mundnich2020tiles}. We recruited participants from the USC Keck Hospital and the USC Norris Cancer Center. The schematic of the study protocol is presented in Figure~\ref{fig:study_protocol}.

\vspace{0.5mm}
\noindent \textbf{Recruitment:}
Hospital employees in these two locations were invited to participate in the study. Recruitment efforts included distributing fliers, presenting the study at staff meetings, posting recruitment information online, and outreach to individuals through emails. Individuals expressing interest in participating were directed to our dedicated study information website, where they completed a screening survey to determine their eligibility. Eligible participants (who work as full-time employees in the hospital) were subsequently invited to schedule an in-person enrollment session.

\vspace{1mm}

\noindent \textbf{Enrollment Session:} During the enrollment session, participants were required to complete a comprehensive set of questionnaires assessing a wide range of information. These surveys collect demographic information and variables related to health and perceived job performance. After completing the surveys, research coordinators provided detailed instructions on using Fitbit sensors throughout the ten-week data collection period. We recommended that participants wear Fitbit sensors both in and out of work and recharge batteries while bathing.

\begin{figure*}[ht] {
    \centering

    \caption{Demographics of the primary working unit in the \textit{TILES-2018 Sleep Benchmark} dataset and \textit{Combined TILES Sleep} dataset (combining the \textit{TILES-2018 Sleep Benchmark} dataset released in this paper with the sleep data in our previous \textit{TILES-2018} dataset).}
    
    \begin{tikzpicture}
        \node[draw=none,fill=none] at (0, 0){\includegraphics[width=0.45\linewidth]{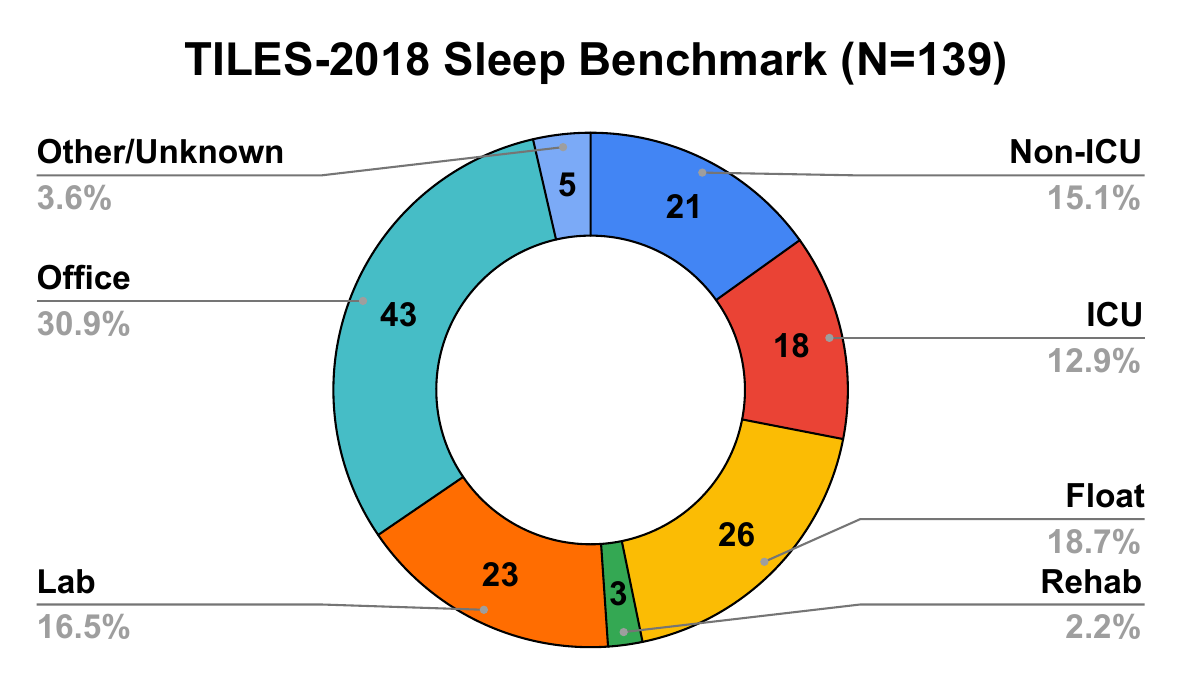}};
        \node[draw=none,fill=none] at (0.5\linewidth, 0){\includegraphics[width=0.45\linewidth]{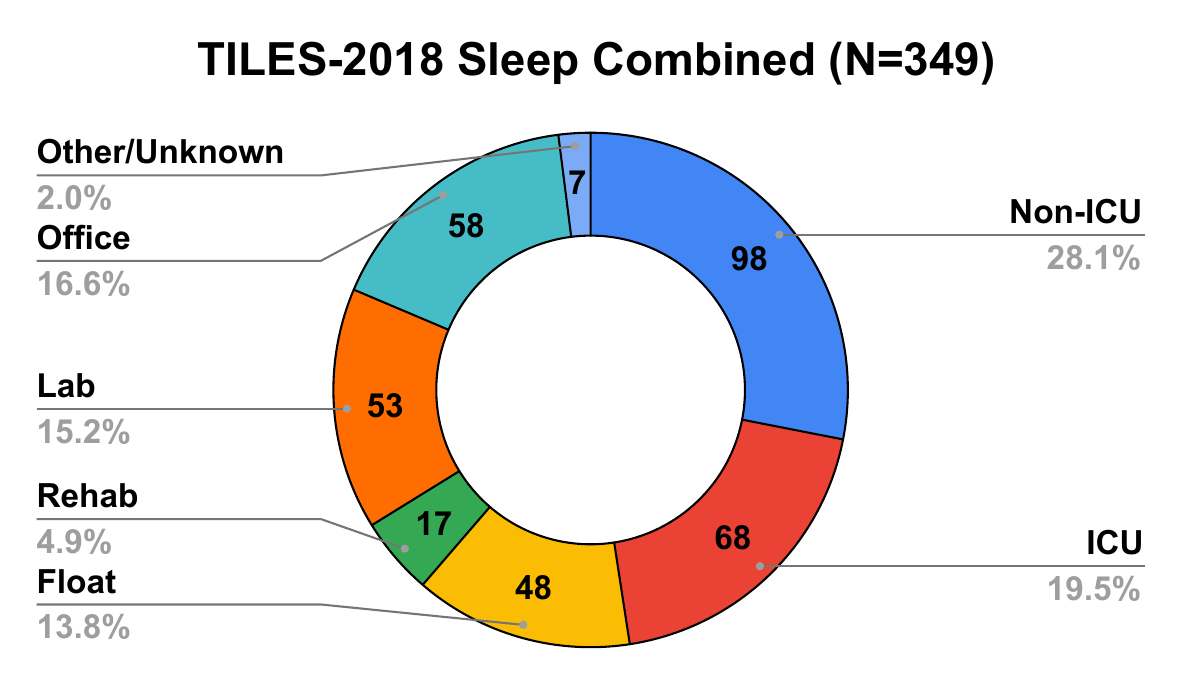}};

        \node[fill=white, align=center, text width=0.45\linewidth] at (0, 2) 
        {\small \textbf{TILES-2018 Sleep Benchmark - N=139} \\ \textbf{(This Work)}};

        \node[fill=white, align=center, text width=0.45\linewidth] at (0.5\linewidth, 2) 
        {\small \textbf{TILES-2018 Sleep Combined - N=349} \\ \textbf{(This Work  and TILES-2018 \cite{mundnich2020tiles})}};
    \end{tikzpicture}
    
    \label{fig:demo_job}
    \vspace{-2.5mm}
} \end{figure*}

\section{Dataset Description}

With the primary focus on studying sleep patterns using wearable sensors, the \textit{TILES-2018 Sleep Benchmark dataset} includes demographics, self-reported sleep surveys, and wearable sleep data. Table \ref{tab:dataset_record} provides an overview of the dataset records, which are described in further detail below.

\label{sec:dataset_description}
\subsection{Demographics}
The \textit{TILES-2018 Sleep Benchmark dataset} includes demographic information about age, sex, primary work shift, and hospital job positions. The demographic data was stored in the \textbf{\texttt{participant-info}} folder with hash-based participant IDs. The primary work shift is classified into day or night shifts.  Age is categorized into groups under 30, over 50, and in 5-year increments in between. Job positions are categorized into Registered Nurses, Certified Nursing Assistants, Monitor Technicians/Therapists, and other unspecified jobs not listed. In addition, we include details about the primary work location: 1) ICU units, 2) non-ICU units, 3) rehabilitation units, 4) float pool (i.e., no assigned unit), 5) laboratories (dialysis, hemo, and other laboratory), 6) office (administration, doctors, and research), and 7) other relevant areas not mentioned above.

\subsection{Self-report Sleep Quality Survey}

To assess sleep quality, we use the Pittsburgh Sleep Quality Index (PSQI) \cite{buysse1989pittsburgh}. This survey includes $29$ items that evaluate seven different dimensions of sleep quality: subjective sleep quality, sleep latency, sleep duration, habitual sleep efficiency, sleep disturbances, use of sleep medication, and daytime dysfunction. Participants responded to each item on a scale from 0 to 3, with higher scores indicating poorer sleep quality. Individual responses are aggregated into a Global PSQI Score, which ranges from 0 to 21. A score of five or above suggests poor overall sleep quality. The raw PSQI responses are stored in the \textbf{\texttt{raw/baseline}} folder, while the aggregated scores are kept in the \textbf{\texttt{scored/baseline}} folder.

\subsection{Wearable Sleep Data}

The Fitbit Charge $2$ captures data about sleep onset, wake times, total sleep duration, and sleep stages (such as light, deep, and REM). Moreover, it tracks continuous heart rate throughout sleep. The specific data entries are listed below:

\vspace{0.5mm}
\noindent \textbf{\texttt{sleep-metadata} folder}:
The sleep-metadata file contains detailed records of sleep metadata for each recorded sleep session, identified by a unique \texttt{sleepId}. Each row includes the start and end time of the sleep, the total sleep time, and whether the sleep is a nap or a main sleep (longest sleep in a day). Moreover, each sleep specifies the time in different sleep phases. The sleep phases are categorized using either classic labels (\texttt{awake, asleep, restless}) or sleep stage labels (\texttt{wake, light sleep, deep sleep, REM sleep}).

\begin{table}
    \footnotesize
    \centering

    \vspace{1mm}

    \begin{tabular}{lcc}

        \toprule
        \multirow{3}{*}{\textbf{}} & 
        \multicolumn{1}{c}{\textbf{TILES-2018}} & 
        \multicolumn{1}{c}{\textbf{TILES-2018}} \\ 

        \multirow{3}{*}{\textbf{}} & 
        \multicolumn{1}{c}{\textbf{Sleep Benchmark}} & 
        \multicolumn{1}{c}{\textbf{Sleep Combined}} \\ 

         & 
        \multicolumn{1}{c}{\textbf{(This work)}} & 
        \multicolumn{1}{c}{\textbf{(This work+TILES-2018 \cite{mundnich2020tiles})}} \\

        \midrule

        \multicolumn{1}{l}{\textbf{Sex}} & & \\
        
        \hspace{0.2cm}{Male} &
        $41$ $(29.5\%)$ &
        $106$ $(30.4\%)$ \\

        \hspace{0.2cm}{Female} &
        $98$ $(70.5\%)$ &
        $243$ $(69.6\%)$ \\
        \midrule
        \multicolumn{1}{l}{\textbf{Primary Shift}} & & \\

        \hspace{0.2cm}{Day Shift} &
        $108$ $(77.7\%)$ &
        $252$ $(72.2\%)$ \\

        \hspace{0.2cm}{Night Shift} &
        $31$ $(22.3\%)$ &
        $97$ $(27.8\%)$ \\
        \midrule
        
        \multicolumn{1}{l}{\textbf{Age}} & & \\

        \hspace{0.2cm}{$<30$ Yr} &
        $18$ $(12.9\%)$ &
        $51$ $(14.6\%)$ \\

        \hspace{0.2cm}{$30-49$ Yr} &
        $91$ $(65.5\%)$ &
        $237$ $(67.9\%)$ \\

        \hspace{0.2cm}{$\geq50$ Yr} &
        $30$ $(21.6\%)$ &
        $61$ $(17.5\%)$ \\

        \bottomrule

    \end{tabular}

    \caption{Demographics of the primary work shift and sex in the \textit{TILES-2018 Sleep Benchmark} and \textit{Combined TILES Sleep} dataset (combining both \textit{TILES-2018 Sleep Benchmark} and \textit{TILES-2018}).}
    \label{tab:demo_shift}

    \vspace{-3.5mm}
\end{table}

\vspace{0.5mm}
\noindent \textbf{\texttt{sleep-heart-rate} folder}: Each file includes the heart rate values (beats per minute) with a corresponding timestamp, measured using the Fitbit's photoplethysmography (PPG) sensor. Each heart rate reading represents the average heart rate values over a one-minute interval. Intervals with missing heart rate values are left empty.

\vspace{0.5mm}
\noindent \textbf{\texttt{sleep-data} folder}: The sleep data file comprises three columns: \texttt{sleepId}, the sleep phase, and the associated duration in seconds. The \texttt{sleepId} and sleep phase are directly linked to the corresponding columns in the \texttt{sleep-metadata} file. The timestamp marks the onset of the sleeping phase.

\begin{table*}
    \small
    \centering
    \scalebox{0.97}{
    \begin{tabular}{p{1.5cm}p{1cm}p{1cm}p{1cm}p{1cm}p{1cm}p{1cm}}

        \toprule
        \multirow{2}{*}{} & 
        \multicolumn{3}{c}{\textbf{TILES-2018 Sleep Benchmark (N = 139)}} & \multicolumn{3}{c}{\textbf{Combined TILES Sleep (N = 349)}} \rule{0pt}{2.25ex} \\  
        
        & \multicolumn{1}{c}{\textbf{Day shift} ($\mu\pm\sigma$)} & 
        \multicolumn{1}{c}{\textbf{Night shift} ($\mu\pm\sigma$)} & \multicolumn{1}{c}{\textbf{\centering p-value}} 
        & \multicolumn{1}{c}{\textbf{Day shift} ($\mu\pm\sigma$)} & 
        \multicolumn{1}{c}{\textbf{Night shift} ($\mu\pm\sigma$)} & \multicolumn{1}{c}{\textbf{\centering p-value}}
        \rule{0pt}{2.25ex} \\
        
        \cmidrule(lr){1-1} \cmidrule(lr){2-4} \cmidrule(lr){5-7}
        
        \multicolumn{1}{l}{\textbf{Total PSQI score}} &
        \multicolumn{1}{c}{$6.20 \pm 3.52$} & 
        \multicolumn{1}{c}{$\mathbf{7.77 \pm 3.89}$} & 
        \multicolumn{1}{c}{$\mathbf{<0.05}$} &
        \multicolumn{1}{c}{$6.10 \pm 3.09$} & 
        \multicolumn{1}{c}{$\mathbf{7.55 \pm 3.64}$} & 
        \multicolumn{1}{c}{$\mathbf{<0.01}$} \rule{0pt}{2.25ex} \\

        \multicolumn{1}{l}{\hspace{0.5cm}\rotatebox[origin=c]{180}{$\Lsh$}Subjective quality} &
        \multicolumn{1}{c}{$1.07$ $\pm$ $0.73$} & 
        \multicolumn{1}{c}{$\mathbf{1.45 \pm 0.71}$} & 
        \multicolumn{1}{c}{$\mathbf{<0.05}$} &
        \multicolumn{1}{c}{$1.05$ $\pm$ $0.67$} & 
        \multicolumn{1}{c}{$\mathbf{1.25 \pm 0.67}$} & 
        \multicolumn{1}{c}{$\mathbf{<0.05}$} \rule{0pt}{2ex} \\

        \multicolumn{1}{l}{\hspace{0.5cm}\rotatebox[origin=c]{180}{$\Lsh$}Sleep latency} &
        \multicolumn{1}{c}{$1.27$ $\pm$ $1.04$} & 
        \multicolumn{1}{c}{$1.48$ $\pm$ $0.98$} & 
        \multicolumn{1}{c}{$0.31$} &
        \multicolumn{1}{c}{$1.31$ $\pm$ $1.00$} & 
        \multicolumn{1}{c}{$1.40$ $\pm$ $0.98$} & 
        \multicolumn{1}{c}{$0.53$} \rule{0pt}{2ex} \\

        \multicolumn{1}{l}{\hspace{0.5cm}\rotatebox[origin=c]{180}{$\Lsh$}Sleep duration} &
        \multicolumn{1}{c}{$1.01$ $\pm$ $0.86$} & 
        \multicolumn{1}{c}{$1.16$ $\pm$ $0.99$} & 
        \multicolumn{1}{c}{$0.37$} &
        \multicolumn{1}{c}{$0.91$ $\pm$ $0.88$} & 
        \multicolumn{1}{c}{$1.05$ $\pm$ $0.95$} & 
        \multicolumn{1}{c}{$0.14$} \rule{0pt}{2.25ex} \\

        \multicolumn{1}{l}{\hspace{0.5cm}\rotatebox[origin=c]{180}{$\Lsh$}Sleep efficiency} &
        \multicolumn{1}{c}{$0.58$ $\pm$ $1.02$} & 
        \multicolumn{1}{c}{${0.87 \pm 1.01}$} & 
        \multicolumn{1}{c}{${0.19}$} &
        \multicolumn{1}{c}{$0.48$ $\pm$ $0.87$} & 
        \multicolumn{1}{c}{$\mathbf{1.01 \pm 1.12}$} & 
        \multicolumn{1}{c}{$\mathbf{<0.01}$} \rule{0pt}{2ex} \\

        \multicolumn{1}{l}{\hspace{0.5cm}\rotatebox[origin=c]{180}{$\Lsh$}Sleep disturbance} &
        \multicolumn{1}{c}{$1.14$ $\pm$ $0.58$} & 
        \multicolumn{1}{c}{$1.26$ $\pm$ $0.62$} & 
        \multicolumn{1}{c}{$0.28$} &
        \multicolumn{1}{c}{$1.20$ $\pm$ $0.57$} & 
        \multicolumn{1}{c}{$1.29$ $\pm$ $0.56$} & 
        \multicolumn{1}{c}{$0.08$} \rule{0pt}{2ex} \\

        \multicolumn{1}{l}{\hspace{0.5cm}\rotatebox[origin=c]{180}{$\Lsh$}Sleep medication} &
        \multicolumn{1}{c}{$0.33$ $\pm$ $0.77$} & 
        \multicolumn{1}{c}{$0.61$ $\pm$ $1.04$} & 
        \multicolumn{1}{c}{$0.14$} &
        \multicolumn{1}{c}{$0.41$ $\pm$ $0.89$} & 
        \multicolumn{1}{c}{$0.59$ $\pm$ $1.02$} & 
        \multicolumn{1}{c}{$0.16$} \rule{0pt}{2ex} \\

        \multicolumn{1}{l}{\hspace{0.5cm}\rotatebox[origin=c]{180}{$\Lsh$}Daytime dysfunction} &
        \multicolumn{1}{c}{$0.80$ $\pm$ $0.84$} & 
        \multicolumn{1}{c}{${0.94 \pm 1.11}$} & 
        \multicolumn{1}{c}{$0.55$} &
        \multicolumn{1}{c}{$0.75$ $\pm$ $0.77$} & 
        \multicolumn{1}{c}{$\mathbf{0.95 \pm 0.93}$} & 
        \multicolumn{1}{c}{$\mathbf{<0.05}$} \rule{0pt}{2ex} \\ 
        \bottomrule

    \end{tabular}
    }
    \caption{Comparing PSQI scores ($0$-$21$) in two distinct working shifts, with a higher score indicating worse sleep quality. The boldface indicates statistical significance at $p<0.05$.}
    \label{tab:psqi_tabel}
\end{table*}

\begin{figure*}[ht] {
    \centering
    \caption{Distribution of main sleep entries per participant and details of main sleep categorized by recording quality.}

    \begin{subtable}[b]{0.54\linewidth}
        \resizebox{\linewidth}{!}{
        \begin{tabular}{lcc}

        \toprule
        \multirow{3}{*}{\textbf{}} & 
        \multicolumn{1}{c}{\textbf{TILES-2018}} & 
        \multicolumn{1}{c}{\textbf{TILES-2018}} \\ 

        \multirow{3}{*}{\textbf{}} & 
        \multicolumn{1}{c}{\textbf{Sleep Benchmark}} & 
        \multicolumn{1}{c}{\textbf{Sleep Combined}} \\ 

         & 
        \multicolumn{1}{c}{\textbf{(This work)}} & 
        \multicolumn{1}{c}{\textbf{(This work+TILES-2018 \cite{mundnich2020tiles})}} \\

        \midrule

        \multicolumn{1}{l}{\textbf{Number of Main Sleeps}} & & \\
        
        \hspace{0.2cm}{$<10$ Main Sleeps} &
        $23$ $(16.5\%)$ &
        $57$ $(16.3\%)$ \\

        \hspace{0.2cm}{$10-29$ Main Sleeps} &
        $25$ $(18.0\%)$ &
        $49$ $(14.0\%)$ \\

        \hspace{0.2cm}{$30-49$ Main Sleeps} &
        $15$ $(10.8\%)$ &
        $60$ $(17.2\%)$ \\

        \hspace{0.2cm}{$\geq 50$ Main Sleeps} &
        $76$ $(54.7\%)$ &
        $183$ $(52.4\%)$ \\

        \bottomrule
        \end{tabular}
        }
        \vspace{10mm}
        \caption{Distribution of main sleep entries per participant in the \textit{TILES-2018 Sleep Benchmark} and \textit{Combined TILES Sleep} dataset. We can observe that most participants (approximately $70\%$) have more than $30$ sleep hours recorded in the study.}
        \label{tab:sleep_distribution}
    \end{subtable}
    \hfill
    \begin{subfigure}[b]{0.43\linewidth}
    \centering
    \includegraphics[width=0.6\linewidth]{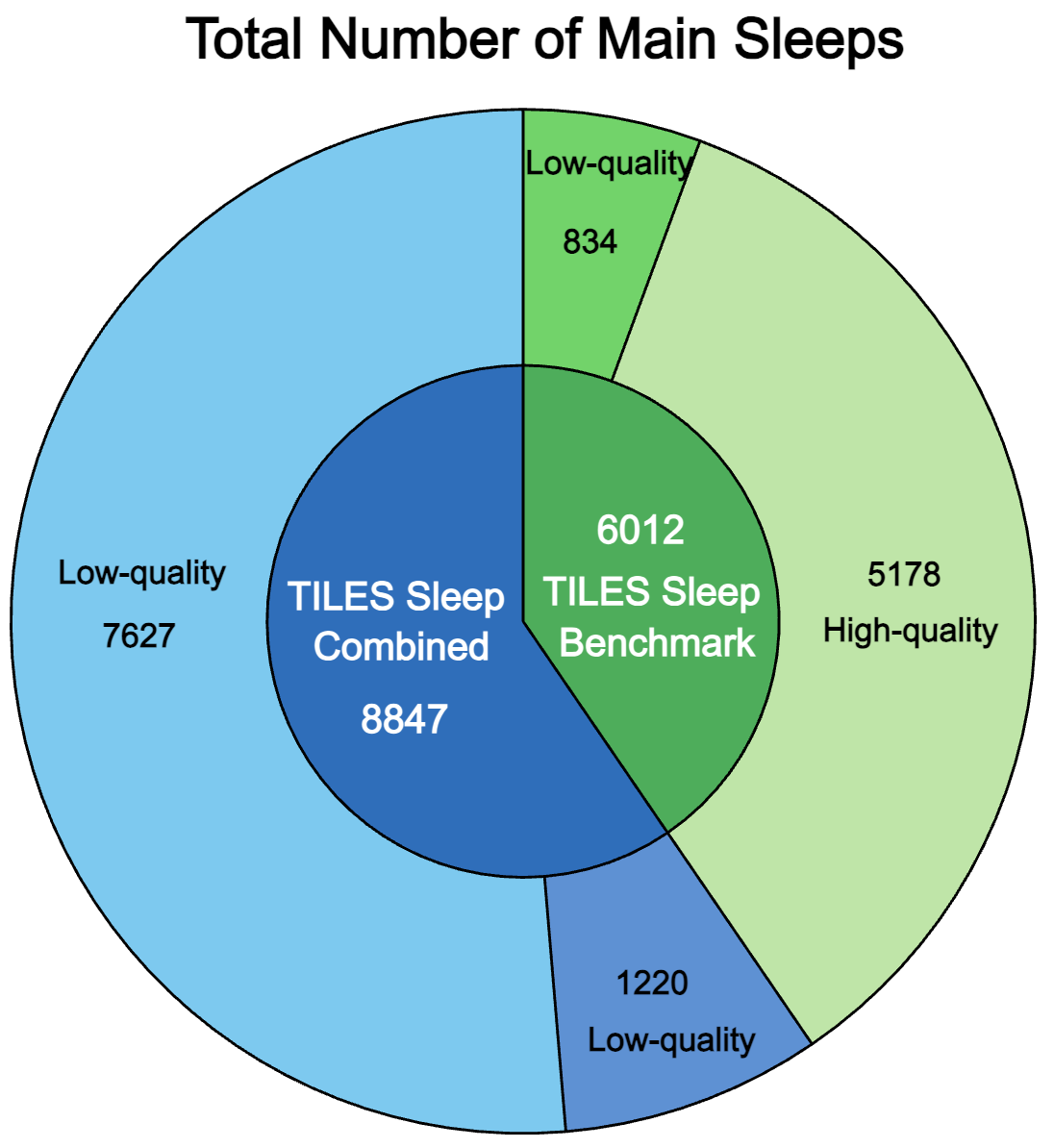}
    \caption{Details of main sleep categorized by recording quality in \textit{TILES-2018 Sleep Benchmark} and \textit{Combined TILES Sleep} datasets. The high-quality sleep data includes information on sleep stages and more than $90\%$ of heart rate data present.}
    \label{fig:sleep_stats}
    \end{subfigure}
    \vspace{-2.5mm}
} \end{figure*}

\section{Dataset Statistics}
\label{sec:dataset}

As the \textit{TILES-2018 Sleep Benchmark} follows the study logistics of the \textit{TILES-2018} dataset \cite{mundnich2020tiles}, we decided to combine sleep data from both datasets for certain analyses. We refer to this combined dataset as the \textit{Combined TILES Sleep} dataset in the following sections.

\subsection{Demographics}

Figure~\ref{fig:demo_job} visualizes the primary working unit within the \textit{TILES-2018 Sleep Benchmark} and the \textit{Combined TILES Sleep} dataset. In the \textit{TILES-2018 Sleep Benchmark}, $139$ participants completed the enrollment session with the PSQI. Of the study participants, $15.1\%$ worked in the non-ICU unit, $12.9\%$ worked in the ICU unit, and $18.7\%$ worked as the float pool. Notably, over $30\%$ of subjects in \textit{TILES-2018 Sleep Benchmark} worked in office settings, including as administrators, doctors, or in research roles. When we combine the \textit{TILES-2018 Sleep Benchmark} data with the data in the \textit{TILES-2018}, the total number of subjects increases to $349$ (from 210 in the original data). Unlike the higher proportion of subjects who worked in offices, a majority of are now from the nursing units in the \textit{Combined TILES Sleep} dataset, with $28.3\%$ working in non-ICU units and $19.4\%$ in ICU units.

Table~\ref{tab:demo_shift} presents the demographic breakdown of participants by sex (male/female) and primary work shift (day shift/night shift). Among these participants, $77.7\%$ ($n=108$) primarily worked the day shift, and $23.3\%$ ($n=31$) worked the night shift. The majority of participants were female ($n=98, 70.5\%$). Of the study participants, $40.3\%$ were registered nurses, and $10.8\%$ were certified nursing assistants. Within the combined \textit{Combined TILES Sleep}, $97$ participants primarily worked the night shift, and $168$ were registered nurses.

\begin{figure*}[ht] {
    \centering
    \caption{Distribution of the sleep onset and wake-up time between different primary working shifts and primary working units.}
    \begin{tikzpicture}
        \node[draw=none,fill=none] at (0, 0){\includegraphics[width=0.265\linewidth]{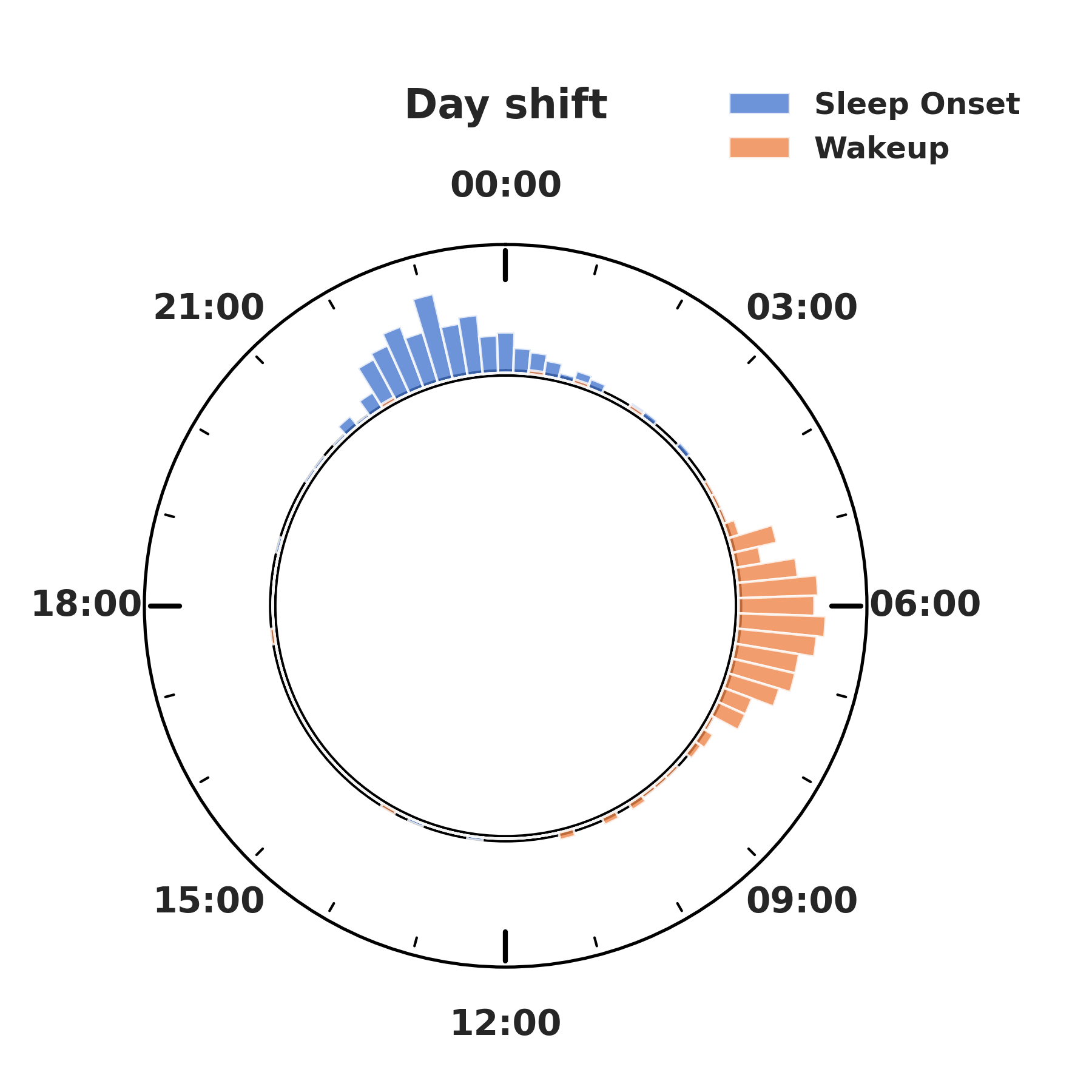}};
        \node[draw=none,fill=none] at (0.25\linewidth, 0){\includegraphics[width=0.265\linewidth]{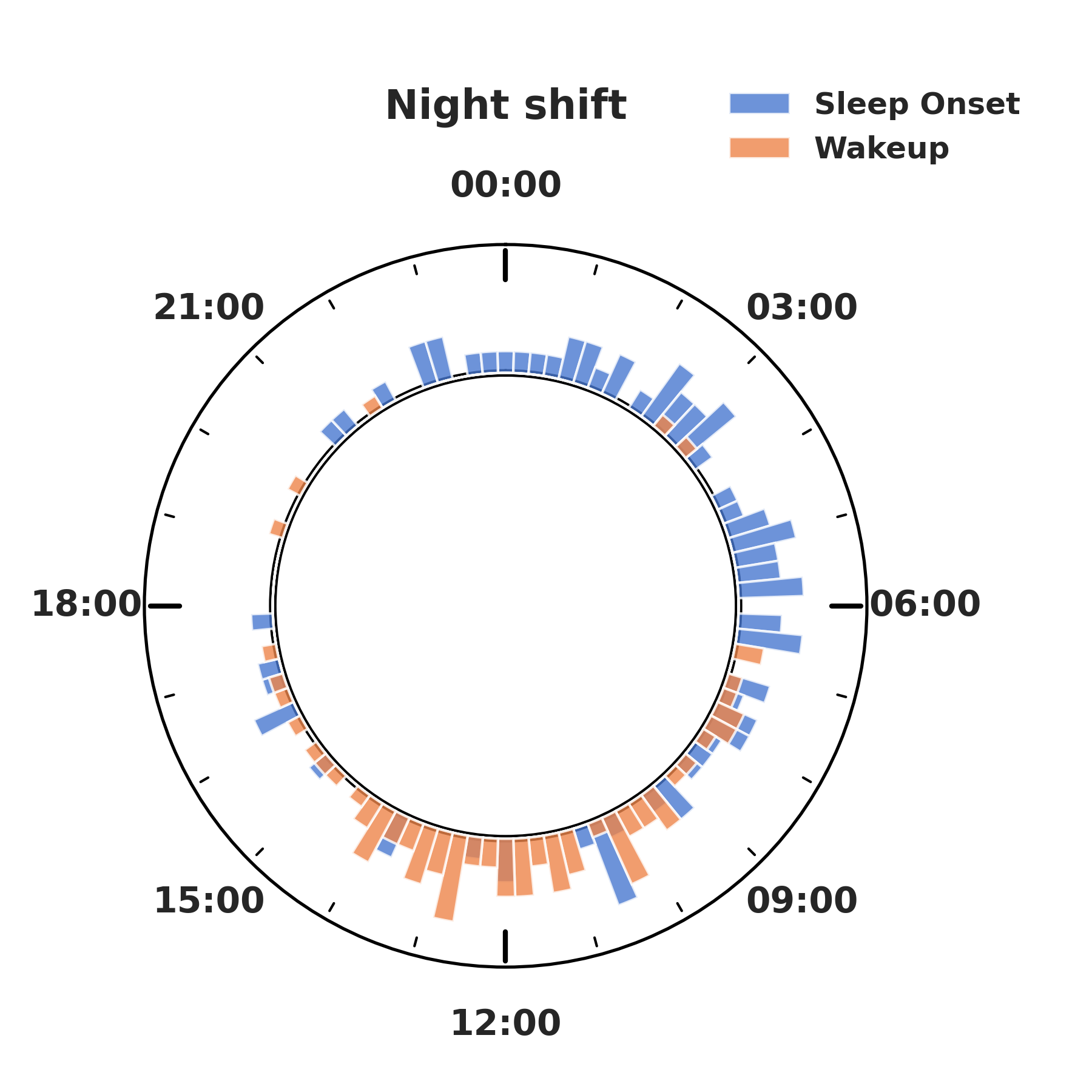}};
        \node[draw=none,fill=none] at (0.5\linewidth, 0){\includegraphics[width=0.265\linewidth]{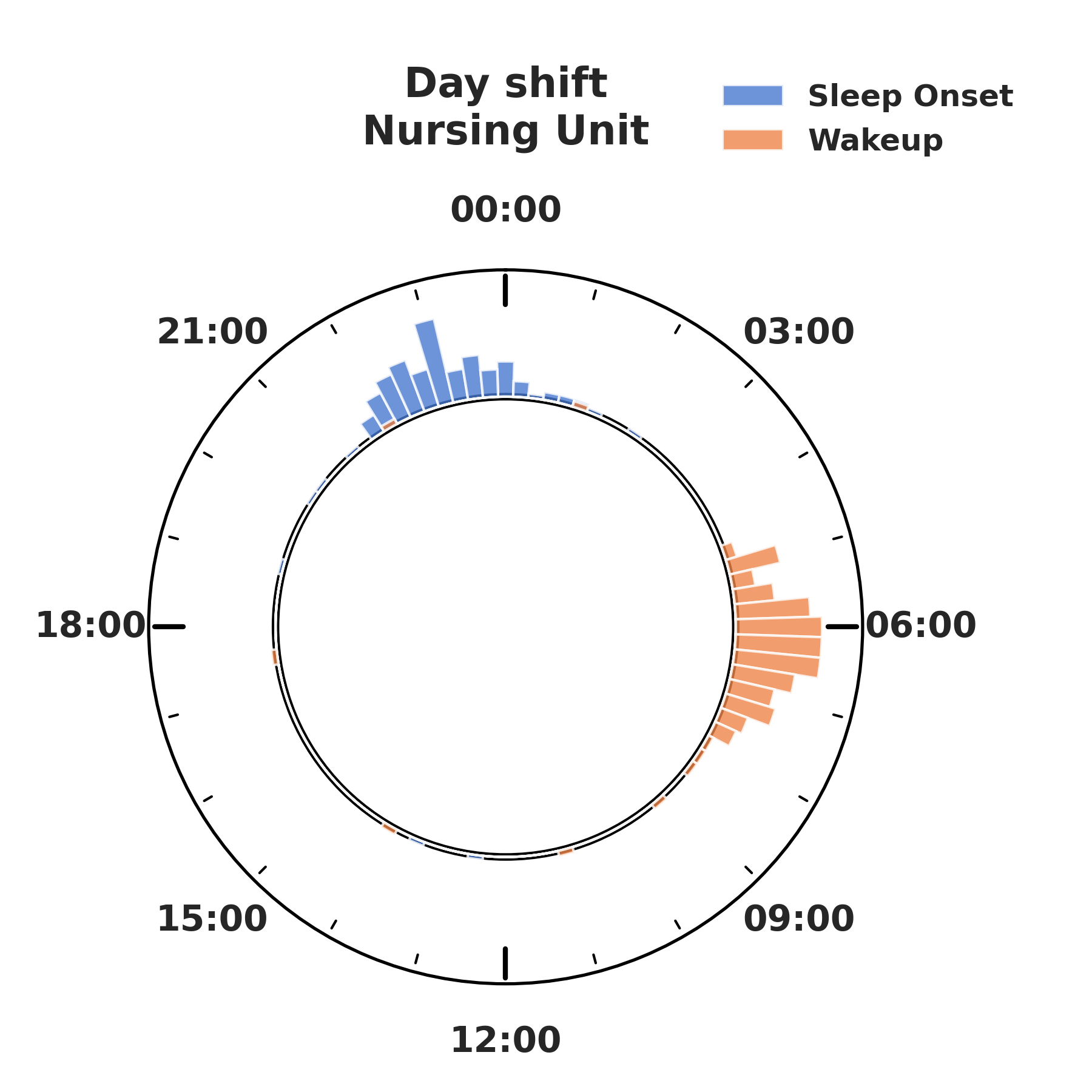}};
        \node[draw=none,fill=none] at (0.75\linewidth, 0){\includegraphics[width=0.265\linewidth]{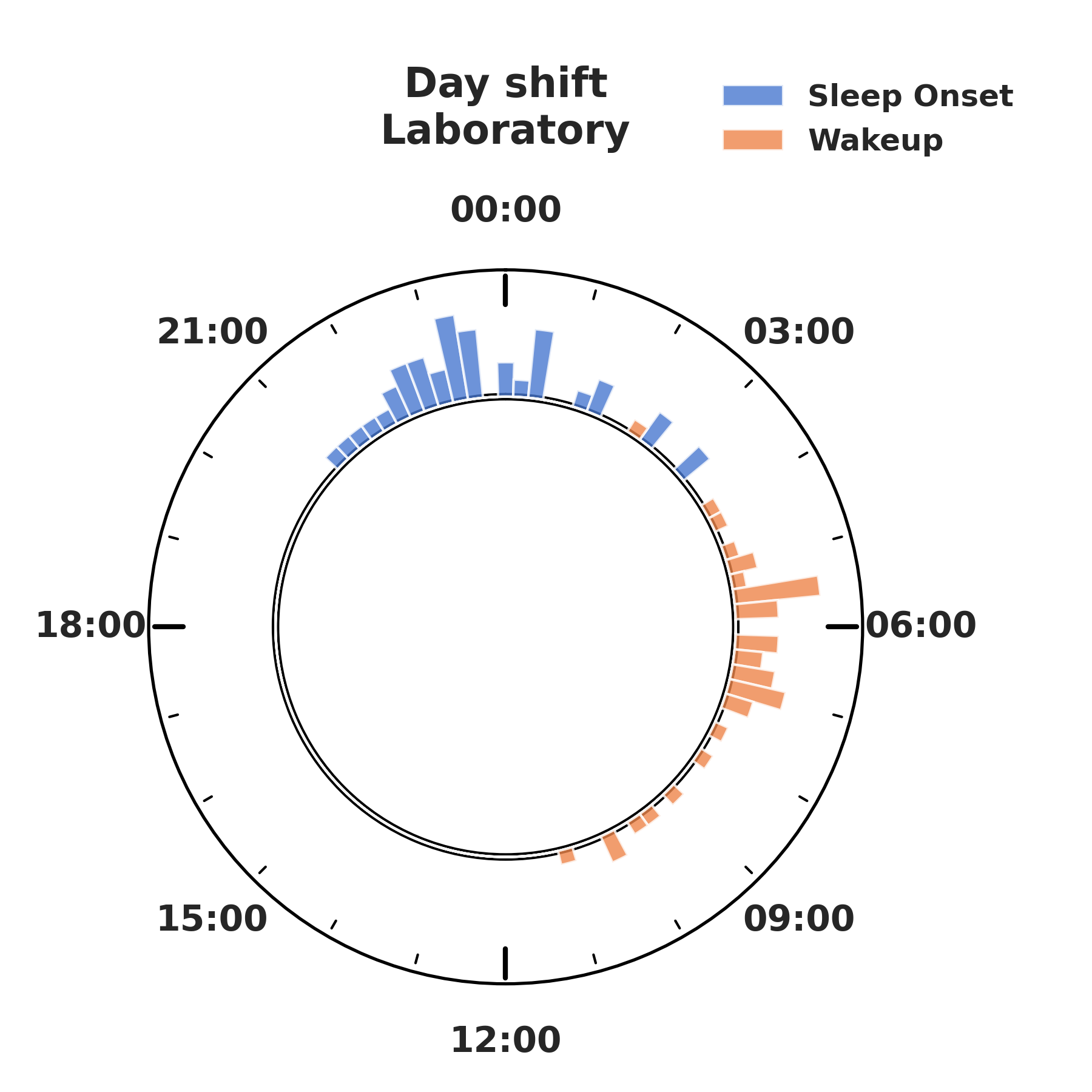}};

    \end{tikzpicture}
    
    \label{fig:sleep_start_end}
    \vspace{-3.5mm}
} \end{figure*}

\subsection{Self-reported Sleep Quality}

Table~\ref{tab:psqi_tabel} summarizes the comparisons between shift work and the subgroups of sleep scores in the \textit{TILES-2018 Sleep Benchmark} and the \textit{Combined TILES Sleep} datasets, respectively. We performed a three-way ANOVA to examine the impact of primary work shift on PSQI scores, accounting for age and sex as additional factors. In the \textit{TILES-2018 Sleep Benchmark} dataset, the total PSQI score is $6.20\pm3.52$ for participants working the day shift and $7.77\pm3.89$ for those working the night shift ($p<0.05$). This difference indicates that night shift workers have overall worse sleep quality than those who work the day shift. We observe a similar pattern of total PSQI score in the \textit{Combined TILES Sleep} dataset.

Table~\ref{tab:psqi_tabel} also shows the scores of different dimensions in the PSQI, which range between 0 and 3. In addition, participants working the night shift reported worse subjective sleep quality than those working the day shift ($p<0.05$). While most PSQI subscales showed no notable differences between shifts in the \textit{TILES-2018 Sleep Benchmark} dataset, there are significant differences between shift schedules and sleep quality in terms of sleep efficiency ($p<0.01$) and daytime dysfunction ($p<0.05$) in the \textit{Combined TILES Sleep} dataset. These findings suggest that night-shift workers tend to spend less time sleeping while in bed and experience greater difficulty focusing on daily tasks.

\begin{table}
    \footnotesize
    \centering

    \vspace{1mm}
    \begin{subtable}{\linewidth}
        
        \begin{tabular}{lcccc}
    
            \toprule
            \multirow{2}{*}{\textbf{}} & 
            \multicolumn{1}{c}{\textbf{Day shift}} & 
            \multicolumn{1}{c}{\textbf{Night shift}} & 
            \multicolumn{1}{c}{\multirow{2}{*}{\textbf{\centering p-val}}} \\ 
            
            & \multicolumn{1}{c}{$\mu\pm\sigma$} 
            & \multicolumn{1}{c}{$\mu\pm\sigma$}
            & \rule{0pt}{1.85ex} \\
            
            \midrule
    
            \multicolumn{1}{l}{\textbf{TILES-2018 Sleep Benchmark}} & & \\
            
            \hspace{0.2cm}\rotatebox[origin=c]{180}{$\Lsh$}{Total Sleep Minutes} &
            $398\pm68$ &
            $352\pm75$ &
            $\mathbf{<0.01^{*}}$ \rule{0pt}{1.85ex} \\
    
            \hspace{0.2cm}\rotatebox[origin=c]{180}{$\Lsh$}{Non-REM Sleep Minutes} &
            $299\pm38$ &
            $291\pm39$ &
            $0.25$ \rule{0pt}{1.85ex} \\
    
            \hspace{0.2cm}\rotatebox[origin=c]{180}{$\Lsh$}{Non-REM Sleep Ratio (\%)} &
            $72.0\pm3.1$ &
            $72.9\pm3.2$ &
            $0.14$ \rule{0pt}{1.85ex} \\
    
            \hspace{0.2cm}\rotatebox[origin=c]{180}{$\Lsh$}{REM Sleep Minutes} &
            $84\pm16$ &
            $78\pm18$ &
            $0.09$ \rule{0pt}{1.85ex} \\
    
            \hspace{0.2cm}\rotatebox[origin=c]{180}{$\Lsh$}{REM Sleep Ratio (\%)} &
            $19.9\pm3.0$ &
            $18.9\pm3.4$ &
            $0.08$ \rule{0pt}{1.85ex} \\
            
            \midrule
            
            \multicolumn{1}{l}{\textbf{Combined TILES Sleep}} & & \\
            
            \hspace{0.2cm}\rotatebox[origin=c]{180}{$\Lsh$}{Total Sleep Minutes} &
            $404\pm69$ &
            $365\pm72$ &
            $\mathbf{<0.01^{*}}$ \rule{0pt}{1.85ex} \\
    
            \hspace{0.2cm}\rotatebox[origin=c]{180}{$\Lsh$}{Non-REM Sleep Minutes} &
            $306\pm39$ &
            $292\pm46$ &
            $\mathbf{<0.01^{*}}$ \rule{0pt}{1.85ex} \\
    
            \hspace{0.2cm}\rotatebox[origin=c]{180}{$\Lsh$}{Non-REM Sleep Ratio (\%)} &
            $72.0\pm3.1$ &
            $73.4\pm3.7$ &
            $\mathbf{<0.01^{*}}$ \rule{0pt}{1.85ex} \\
    
            \hspace{0.2cm}\rotatebox[origin=c]{180}{$\Lsh$}{REM Sleep Minutes} &
            $85\pm17$ &
            $75\pm19$ &
            $\mathbf{<0.01^{*}}$ \rule{0pt}{1.85ex} \\
    
            \hspace{0.2cm}\rotatebox[origin=c]{180}{$\Lsh$}{REM Sleep Ratio (\%)} &
            $19.9\pm3.3$ &
            $18.4\pm3.8$ &
            $\mathbf{<0.01^{*}}$ \rule{0pt}{1.85ex} \\
            
            \bottomrule
    
        \end{tabular}

    \end{subtable}
    \label{tab:sleep_stats}

    \caption{Comparing sleep patterns between day and night shift participants. Statistical significance was noted as $\mathbf{p^{*}<0.01}$. Only participants with more than $10$ sleep recordings are kept for the analysis.}
    \vspace{-3.5mm}
\end{table}

\subsection{Sleep Data}

Table~\ref{tab:sleep_distribution} shows the distribution of unique main sleep entries per participant in the \textit{TILES-2018 Sleep Benchmark} dataset. Specifically, more than half of the participants ($N = 76$) recorded over 50 main sleep entries during the 10-week study period, whereas only $N = 23$ ($16.5\%$) had fewer than 10 entries. A similar pattern is observed in the \textit{Combined TILES Sleep} dataset. These distributions suggest a high level of participant compliance in wearing Fitbit throughout the study.

Figure~\ref{fig:sleep_stats} shows the total number of main sleep records in the \textit{TILES-2018 Sleep Benchmark} dataset. For our subsequent sleep analysis and modeling, we selected participants with more than ten main sleep entries, resulting in a total of $6,012$ unique main sleep records. Among these, $86.1\%$ ($n=5,178$) entries are considered \textit{high-quality}, which includes complete sleep stage information and over $90\%$ continuous heart rate values. In contrast, the remaining entries use classic labels (asleep, awake, and restless) or have a higher rate of missing heart rate values. In the \textit{Combined TILES} dataset, this leads to a total of $12,805$ in high-quality sleep entries, making this combined dataset a valuable data source for deep learning research on wearable data.

\begin{figure}[ht] {
    \centering
    \caption{The transition probability graph of sleep stages among nurses working day and night shift schedules. The probability value is calculated by averaging the transition probability graphs across participants. The probability of less than $1\%$ is not shown in the graph. The text and arrow in blue indicate significant differences in sleep stage transition probabilities.}

    \begin{subfigure}{0.6\linewidth}
    \includegraphics[width=\linewidth]{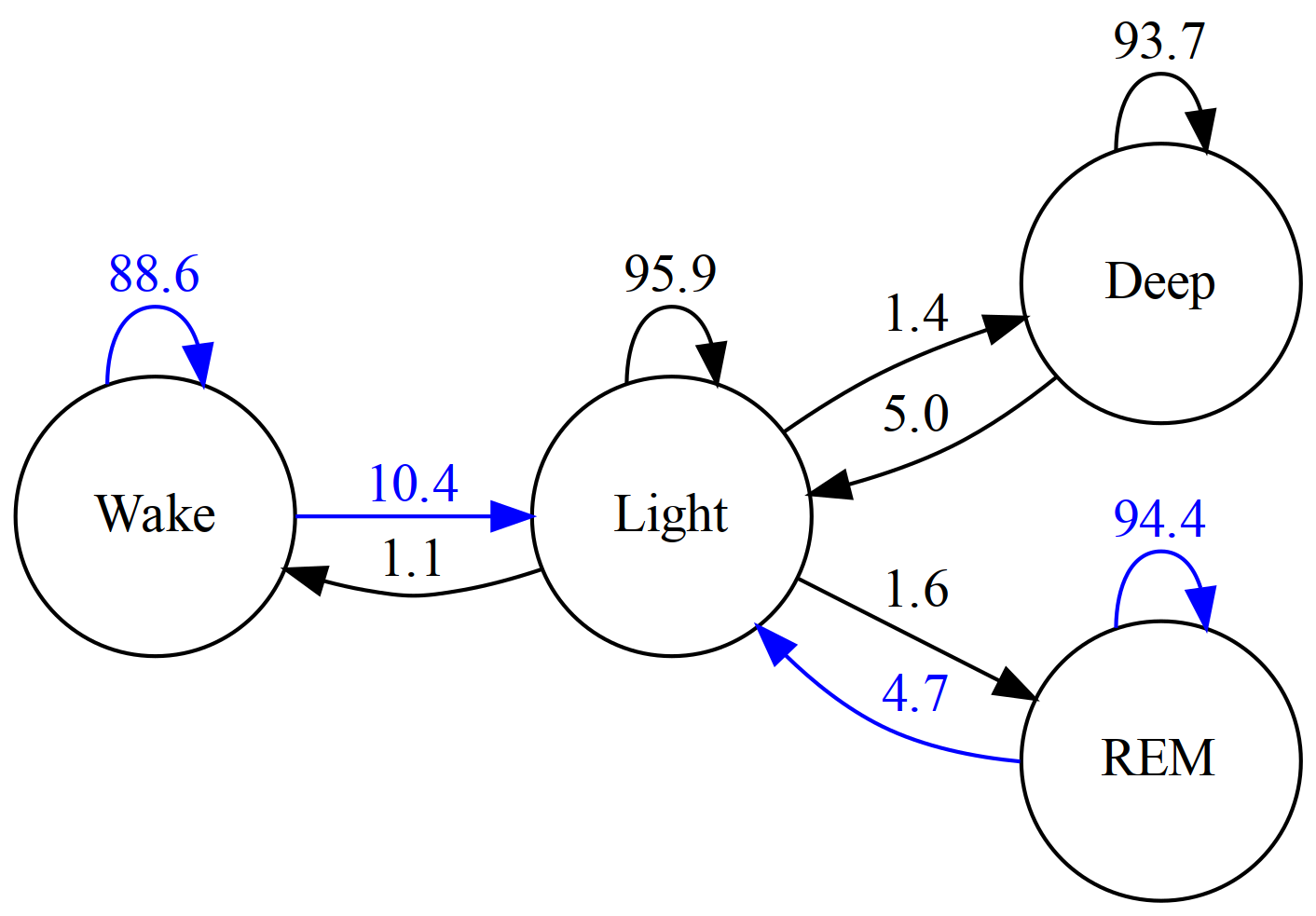}
    \vspace{-3mm}
    \caption{Day-shift Nurses.}
    \end{subfigure}

    \begin{subfigure}{0.6\linewidth}
    \includegraphics[width=\linewidth]{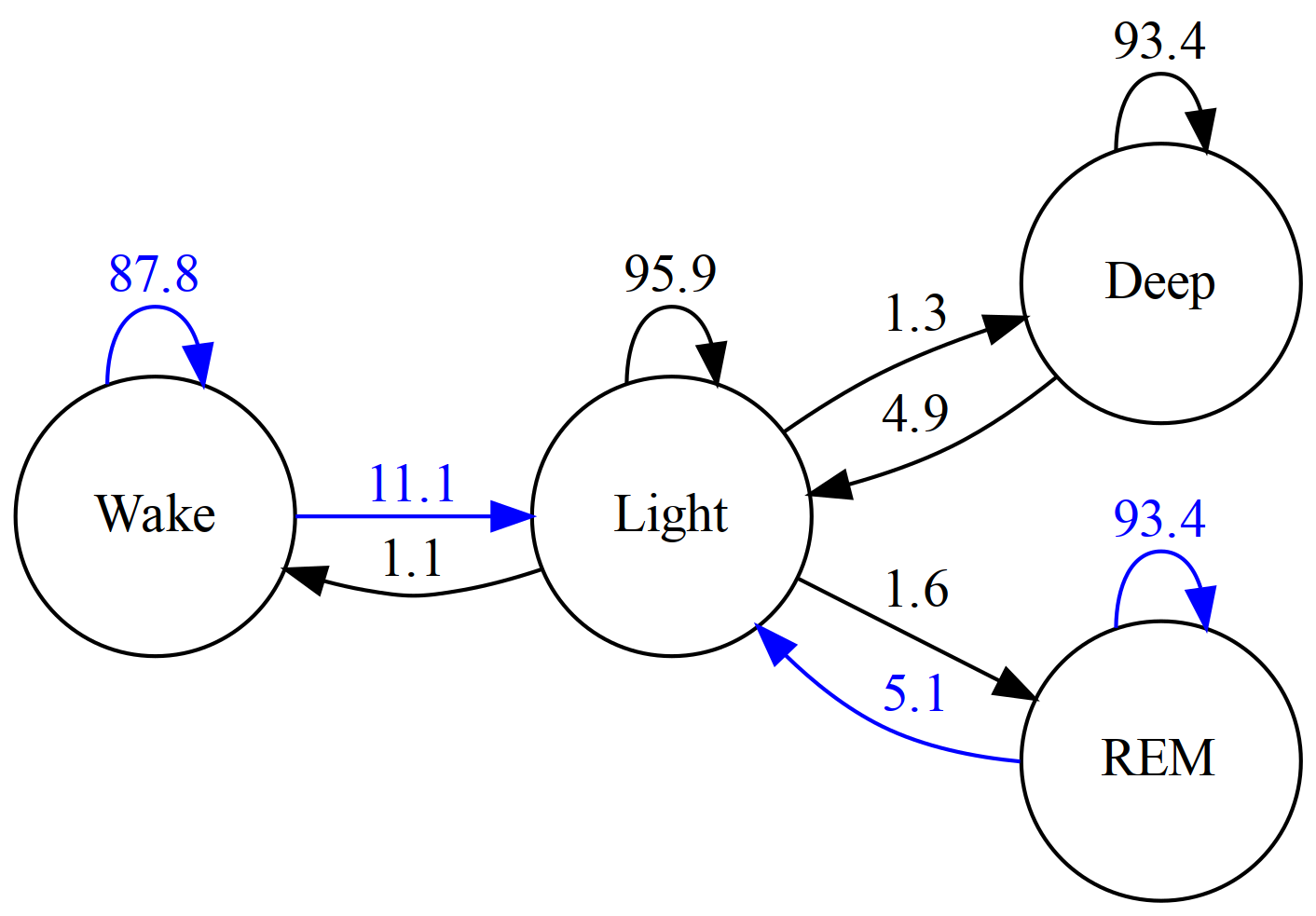}
    \vspace{-3mm}
    \caption{Night-shift Nurses.}
    \end{subfigure}
    
    \label{fig:sleep_stage_transition}
    \vspace{-3.5mm}
} \end{figure}

\begin{figure*}
    \centering
    
    \begin{subfigure}{\linewidth}
        \includegraphics[width=\linewidth]{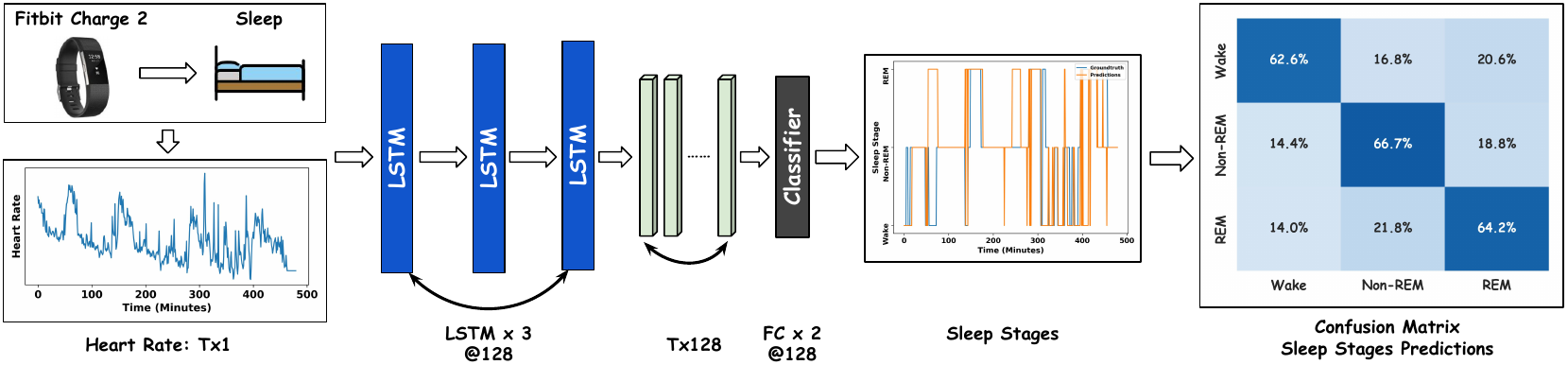}

        \vspace{0.5mm}
        \caption{Sleep stage classification pipeline and classification results in the confusion matrix. The model consists of a 3-layer LSTM and an MLP-based classifier. \textit{TILES-2018} is used for training and validation, \textit{TILES-2018 Sleep Benchmark} is applied for evaluation.}
        \label{tab:sleep_stage_modeling}
    \end{subfigure}

    \vspace{7mm}
    
    \begin{subtable}{\linewidth}
        \resizebox{\linewidth}{!}{
        \begin{tabular}{lcccccccccccc}
    
            \toprule
            & \multicolumn{1}{c}{\textbf{All Participants}} & \multicolumn{1}{c}{\textbf{Day-shift}} & \multicolumn{1}{c}{\textbf{Night-shift}} & \multirow{2}{*}{\textbf{p-val}} &
            \multicolumn{1}{c}{\textbf{Age$\mathbf{\leq40}$}} & 
            \multicolumn{1}{c}{\textbf{Age$\mathbf{>40}$}} &
            \multirow{2}{*}{\textbf{p-val}} \\ 
    
            & \textbf{(n = 116)} & 
            \textbf{(n = 88)} & \textbf{(n = 28)} & &
            \textbf{(n = 76)} & \textbf{(n = 40)} & \rule{0pt}{1.65ex} \\

            \cmidrule(lr){1-1} \cmidrule(lr){2-2} \cmidrule(lr){3-5} \cmidrule(lr){6-8}
    
            \textbf{REM Sleep Classification (Wake/Non-REM/REM)} & & \rule{0pt}{1.65ex} \\
            
            \hspace{0.5cm}{SleepNet - 3-Layer LSTM} & 
            $\mathbf{0.574}$ & 
            $0.574$ & 
            $0.572$ & 
            $0.14$ & 
            $0.577$ & 
            $0.566$ & 
            $0.14$ 
            \rule{0pt}{1.65ex} \\
    
            \hspace{0.5cm}{Single-Layer TimesNet} & ${0.570}$ & 
            $0.573$ & 
            $0.564$ & 
            $0.27$ & 
            $0.574$ & 
            $0.565$ & 
            $0.18$ 
            \rule{0pt}{1.65ex} \\
            
            \cmidrule(lr){1-1} \cmidrule(lr){2-2} \cmidrule(lr){3-5} \cmidrule(lr){6-8}
            
            \textbf{Detailed Sleep Classification (Wake/Light/Deep/REM)} & & \rule{0pt}{1.65ex} \\
            
            \hspace{0.5cm}{SleepNet - 3-Layer LSTM} & 
            $\mathbf{0.529}$ & 
            $0.527$ & 
            $0.531$ & 
            $0.96$ & 
            $0.537$ & 
            $0.511$ & 
            $<0.01$
            \rule{0pt}{1.85ex} \\
            
            \hspace{0.5cm}{Single-Layer TimesNet} & $0.500$ & 
            $0.499$ & 
            $0.506$ & 
            $0.78$ & 
            $0.509$ & 
            $0.486$ & 
            $<0.01$
            \rule{0pt}{1.65ex} \\
            \bottomrule
        
        \end{tabular}
        }
        \vspace{1mm}
        \caption{Benchmark results (averaged macro F1) of sleep stage classification on the \textit{TILES-2018 Sleep Benchmark} dataset.}
        \label{tab:sleep_stage_results}
    \end{subtable}

    \vspace{2mm}
    \caption{Sleep stage classification benchmark results on \textit{TILES-2018 Sleep Benchmark} dataset. We use the sleep data from \textit{TILES-2018} as the training data and report the performance on the \textit{TILES-2018 Sleep Benchmark} dataset. We only include participants with more than $10$ sleep recordings in this experiment.}
    \vspace{-1.5mm}
    \label{tab:sleep_stage_ml}
\end{figure*}

\section{Sleep Behavior Analysis}
\label{sec:usage}

To highlight the opportunities that our sleep dataset provides to the research community, we first use these data to conduct comparisons of sleep patterns.

\subsection{Comparing Sleep Durations Based on Primary Shift}

We first compare sleep patterns between participants who work primarily day shifts and those who work night shifts. Our analysis uses data from the \textit{TILES-2018 Sleep Benchmark} and the combined \textit{Combined TILES Sleep}. We apply a three-way ANOVA to assess the impact of primary working shifts on sleep patterns, controlling for sex and age as additional factors. The analysis focuses on total sleep minutes, non-REM (Light and Deep) sleep minutes, non-REM sleep ratio, REM sleep minutes, and REM sleep ratio. The findings show that hospital employees who work day shifts consistently have more total sleep minutes than those on night shifts ($p<0.01$). Moreover, the comparisons from the \textit{Combined TILES Sleep} dataset imply that day shift workers have higher REM sleep minutes and REM sleep ratios than night shift workers. However, these differences are marginal in the \textit{TILES-2018 Sleep Benchmark} dataset, indicating the importance of combining more relevant data for a robust analysis of sleep patterns. Our results highlight the value of this combined dataset in sleep science studies related to shift schedules.

\subsection{Sleep Onset and Wake-up Time Comparisons}

We additionally compare the sleep onset and wake-up times between participants with different primary shifts and primary units, as shown in Figure~\ref{fig:sleep_start_end}. The comparisons show that participants who primarily work day shifts have more regular sleep schedules compared to those who work night shifts. Notably, sleep onset and wake-up times for night-shift workers span different times of the day.
Moreover, the average standard deviation of the sleep onset time within a subject is 2.0 hours for the day-shift subjects and 5.6 hours for night-shift subjects ($p<0.01$), respectively. Similarly, the average standard deviation of the wake-up time is 2.1 hours for day-shift subjects and 4.3 hours for night-shift subjects ($p<0.01$), respectively. When comparing the sleep schedules of day-shift participants across different primary units, we found that individuals working in laboratory settings tend to have later sleep onset times and larger variability in both sleep onset and wake-up times, compared to those working in nursing units.

\subsection{Sleep Stage Transition Comparisons}

Finally, we compare sleep stage transition probabilities between participants who primarily work day and night shifts, as shown in Figure~\ref{fig:sleep_stage_transition}. Consistent with our earlier analyses, we perform a three-way ANOVA to evaluate the effect of primary work shift on sleep stage transitions, considering sex and age as additional variables. Our findings reveal that night-shift nurses are significantly more likely to transition from REM sleep to light sleep than day-shift nurses ($p<0.01$). In contrast, day-shift nurses have a higher likelihood of remaining in REM sleep ($p < 0.01$). Additionally, night-shift nurses demonstrate an increased probability of transitioning from the wake state to light sleep stages relative to day-shift nurses (day-shift nurse: $10.4\%$, night-shift nurse: $11.1\%$, $p<0.05$). These results suggest that night-shift work may be more likely to disrupt the continuity of restorative sleep stages, such as REM sleep. This behavior is likely due to circadian misalignment and inconsistent sleep-wake schedules, which are commonly experienced by night-shift workers.

\section{Machine Learning Benchmark on Sleep Data}
\label{sec:sleep_benchmark}

In addition to the analysis of sleep patterns, we present several machine learning benchmarks that use the \textit{TILES-2018 Sleep Benchmark} data as the evaluation set and our previous \textit{TILES-2018} data as the training source. Specifically, we perform sleep stage classifications with heart rate data, demographics classification, and automated prediction of self-report sleep survey outcomes.

\subsection{Benchmarking Sleep Stage Classification}

We first conducted an ML experiment to predict sleep stages from continuous heart rate values. Using sleep data from the \textit{TILES-2018} dataset for training and validation, we evaluated our model on the \textit{TILES-2018 Sleep Benchmark} dataset. We present two benchmark models involving sequence-to-sequence sleep stage labeling: LSTM \cite{hochreiter1997long} and TimesNet \cite{wu2022timesnet}. The first model is a SleepNet-like model with a 3-layer LSTM of hidden size 128, while the second model is a single-layer TimesNet block with a hidden size of 64. Both models utilize an MLP-based classifier for classification. We classify sleep stage prediction tasks into predicting REM sleep or detailed sleep stage classifications. The detailed sleep stage classification predicts wake, light sleep, deep sleep, and REM sleep. In contrast, the REM sleep classification predicts wake, non-REM sleep (which combines light and deep sleep), and REM sleep. 

Table~\ref{tab:sleep_stage_modeling} shows the structure and flow of the LSTM model, and the TimesNet model uses one TimesNet layer with 64 nodes instead of the 3 LSTM layers. The results in Table~\ref{tab:sleep_stage_results} reveal that a simple LSTM-based model yields competitive classification results compared to TimesNet, achieving F1 scores of 0.585 for REM sleep prediction and 0.543 for detailed sleep stage prediction. Additional modeling efforts, such as fine-tuning a pre-trained sleep data model or using a foundation time series model, may further improve the performance of the LSTM model.

\begin{table}
    \small
    \centering
    
    \vspace{1mm}
    \begin{tabular}{lcccc}

        \toprule
         & \textbf{F1} & \textbf{ROC-AUC} & \textbf{Acc} \\ 
        \cmidrule(lr){1-1} \cmidrule(lr){2-4}
        
        \textbf{SleepNet - 3-Layer LSTM} & & & \rule{0pt}{1.65ex} \\

        \multicolumn{1}{l}{\hspace{0.2cm}\rotatebox[origin=c]{180}{$\Lsh$}Raw Sleep} &
        $\mathbf{0.604}$ & 0.655 & 63.6 \\

        \multicolumn{1}{l}{\hspace{0.2cm}\rotatebox[origin=c]{180}{$\Lsh$}Raw Sleep + Demographics} &
        0.471 & 0.475 & 56.1 \\

        \textbf{Single-Layer TimesNet} & & & \rule{0pt}{1.65ex} \\

        \multicolumn{1}{l}{\hspace{0.2cm}\rotatebox[origin=c]{180}{$\Lsh$}Raw Sleep} &
        0.561 & 0.613 & $\mathbf{67.3}$ \\

        \multicolumn{1}{l}{\hspace{0.2cm}\rotatebox[origin=c]{180}{$\Lsh$}Raw Sleep + Demographics} &
        0.573 & 0.584 & 59.8 \\
        
        \textbf{Random Forest} & & & \rule{0pt}{1.65ex} \\

        \multicolumn{1}{l}{\hspace{0.2cm}\rotatebox[origin=c]{180}{$\Lsh$}Sleep Pattern} &
        0.521 & 0.554 & 59.1 \\

        \multicolumn{1}{l}{\hspace{0.2cm}\rotatebox[origin=c]{180}{$\Lsh$}Sleep Pattern + Demographics} &
        0.410 & 0.562 & 63.5 \\

        \textbf{Zero-shot BioLLMs} & & & \rule{0pt}{1.65ex} \\
        
        {\hspace{0.2cm}\rotatebox[origin=c]{180}{$\Lsh$}BioMistral-7B} & $0.284$ & - & $39.7$ \rule{0pt}{1.65ex} \\
        
        {\hspace{0.2cm}\rotatebox[origin=c]{180}{$\Lsh$}BioMistral-7B-DARE} & $0.284$ & - & $39.7$ \rule{0pt}{1.65ex} \\

        \cmidrule(lr){1-1} \cmidrule(lr){2-4}
        
        \textbf{Ensemble} & & & \rule{0pt}{1.65ex} \\
        
        {\hspace{0.2cm}\rotatebox[origin=c]{180}{$\Lsh$}SleepNet + TimesNet} & 0.576 & $\mathbf{0.665}$ & $64.5$ \rule{0pt}{1.65ex} \\

        \bottomrule

    \end{tabular}
    
    \caption{Benchmark results (macro F1, ROC-AUC, and accuracy) on predicting self-reported PSQI scores. We binarize the PSQI score by a threshold of $7$.}
    \label{tab:ml_result}

    \vspace{-1.5mm}
\end{table}

\subsection{Benchmarking Self-report PSQI Prediction}

We use sleep features derived from Fitbit to predict self-reported sleep quality surveys (PSQI). Similar to sleep stage classification, we apply the sleep data from the \textit{TILES-2018} dataset for training and validation and samples from the \textit{TILES-2018 Sleep Benchmark} dataset for evaluation. We binarize the PSQI score with a cutoff score of 7, where a global PSQI score of 7 or greater is often used to indicate poor sleep quality \cite{buysse1989pittsburgh}. The baselines used in this paper are below:

\begin{itemize}[leftmargin=*]
    \vspace{0.5mm}
    \item \textbf{Random Forest Classifier.}  The included sleep patterns to the model are average total sleep time, wake time, light sleep time, deep sleep time, REM sleep time, REM sleep percentage, and the standard deviation of total sleep time. In addition to these sleep-related features, we integrate demographic variables such as age, sex, and primary work shift in the experiment.
    
    \vspace{0.5mm}
    \item \textbf{Large Language Model Prompts.} Given the recent advances in Large Language Models (LLMs), researchers have explored using language models to predict health variables. Inspired by Health-LLM, we perform zero-shot experiments with the BioMistral model families \cite{labrak2024biomistral} to predict self-reported sleep quality. Our prompt construction follows the example in Health-LLM \cite{kim2024health}, incorporating both user demographics and health-related context, such as sleep patterns.
    
    \vspace{0.5mm}
    \item \textbf{Fine-tuned Neural Networks.} We evaluated the performance of the SleepNet and TimesNet models in predicting self-reported sleep quality. During training, we randomly sampled $5$ sleep recordings per participant and assigned the corresponding self-reported PSQI label. This sampling process was repeated 50 times for each participant to ensure robustness. We perform the average pooling from the SleepNet and TimesNet output and feed the embeddings to the classifier.
    Moreover, we consider concatenating demographics to the sleep embeddings before feeding into the classifier.
    We aggregated the predicted probabilities across the sleep recordings during inference to generate the prediction.
    
\end{itemize}

Table~\ref{tab:ml_result} presents a comparison of model performance, measured by macro-F1, AUC-ROC, and accuracy, across the Random Forest classifier, zero-shot large language models (LLMs), SleepNet, TimesNet, and an ensemble of SleepNet and TimesNet. The results show that the traditional tree-based model outperforms zero-shot LLMs in predicting self-reported sleep quality. However, deep time-series models, SleepNet and TimesNet, outperform tree-based models by a large margin measured by F1 and ROC-AUC. Surprisingly, incorporating demographic variables (age, sex, and primary work shift) did not yield performance gains across any model. Notably, the ensemble of SleepNet and TimesNet achieves the highest AUC-ROC of $0.665$.

\begin{table}
    \small
    \centering
    
    \vspace{1mm}
    \begin{tabular}{lcccc}

        \toprule
         & \textbf{F1} & \textbf{ROC-AUC} & \textbf{Acc} \\ 
        \cmidrule(lr){1-1} \cmidrule(lr){2-4}
        
        \textbf{Age [$\mathbf{\leq}$40 Yr $\mathbf{|}$ $>$40 Yr]} & & & \rule{0pt}{1.65ex} \\

        \multicolumn{1}{l}{\hspace{0.2cm}\rotatebox[origin=c]{180}{$\Lsh$}Random Forest} &
        0.417 & $\mathbf{0.725}$ & $\mathbf{65.2}$ \\

        \multicolumn{1}{l}{\hspace{0.2cm}\rotatebox[origin=c]{180}{$\Lsh$}SleepNet - 3-Layer LSTM} &
        0.491 & 0.484 & 48.4 \\

        \multicolumn{1}{l}{\hspace{0.2cm}\rotatebox[origin=c]{180}{$\Lsh$}Single-Layer TimesNet} &
        0.570 & 0.632 & 57.0 \\

        \multicolumn{1}{l}{\hspace{0.2cm}\rotatebox[origin=c]{180}{$\Lsh$}SleepNet + TimesNet} &
        $\mathbf{0.587}$ & 0.638 & 58.9 \\
        \cmidrule(lr){1-1} \cmidrule(lr){2-4}
        
        \textbf{Sex [Male $\mathbf{|}$ Female]} & & & \rule{0pt}{1.65ex} \\

        \multicolumn{1}{l}{\hspace{0.2cm}\rotatebox[origin=c]{180}{$\Lsh$}Random Forest} &
        0.500 & 0.631 & 67.0 \\

        \multicolumn{1}{l}{\hspace{0.2cm}\rotatebox[origin=c]{180}{$\Lsh$}SleepNet - 3-Layer LSTM} &
        0.541 & 0.671 & 57.0 \\

        \multicolumn{1}{l}{\hspace{0.2cm}\rotatebox[origin=c]{180}{$\Lsh$}Single-Layer TimesNet} &
        $\mathbf{0.614}$ & 0.642 & $\mathbf{69.2}$ \\

        \multicolumn{1}{l}{\hspace{0.2cm}\rotatebox[origin=c]{180}{$\Lsh$}SleepNet + TimesNet} &
        0.556 & $\mathbf{0.671}$ & 58.9 \\
        \cmidrule(lr){1-1} \cmidrule(lr){2-4}

        \textbf{Shift [Day-shift $\mathbf{|}$ Night-shift]} & & & \rule{0pt}{1.65ex} \\

        \multicolumn{1}{l}{\hspace{0.2cm}\rotatebox[origin=c]{180}{$\Lsh$}Random Forest} &
        $\mathbf{0.631}$ & 0.754 & $\mathbf{72.2}$ \\

        \multicolumn{1}{l}{\hspace{0.2cm}\rotatebox[origin=c]{180}{$\Lsh$}SleepNet - 3-Layer LSTM} &
        0.545 & 0.740 & 55.1 \\

        \multicolumn{1}{l}{\hspace{0.2cm}\rotatebox[origin=c]{180}{$\Lsh$}Single-Layer TimesNet} &
        0.592 & 0.752 & 60.7 \\

        \multicolumn{1}{l}{\hspace{0.2cm}\rotatebox[origin=c]{180}{$\Lsh$}SleepNet + TimesNet} &
        0.559 & $\mathbf{0.767}$ & 57.0 \\
        
        \bottomrule

    \end{tabular}
    \label{tab:ml_demographics}

    \caption{Benchmark results (macro F1, ROC-AUC, and accuracy) on predicting demographics. We binarize age data using a threshold of $40$.}
    
    \vspace{-1.5mm}
\end{table}

\begin{figure*}[ht] {
    \centering
    \caption{Feature importance in predicting demographics using the sleep-related features with Random Forest models. The ROC-AUC scores in predicting age, sex, and shift are 0.725, 0.631, and 0.754, respectively.}

    \begin{tikzpicture}
        \node[draw=none,fill=none] at (0, 0){\includegraphics[width=0.3\linewidth]{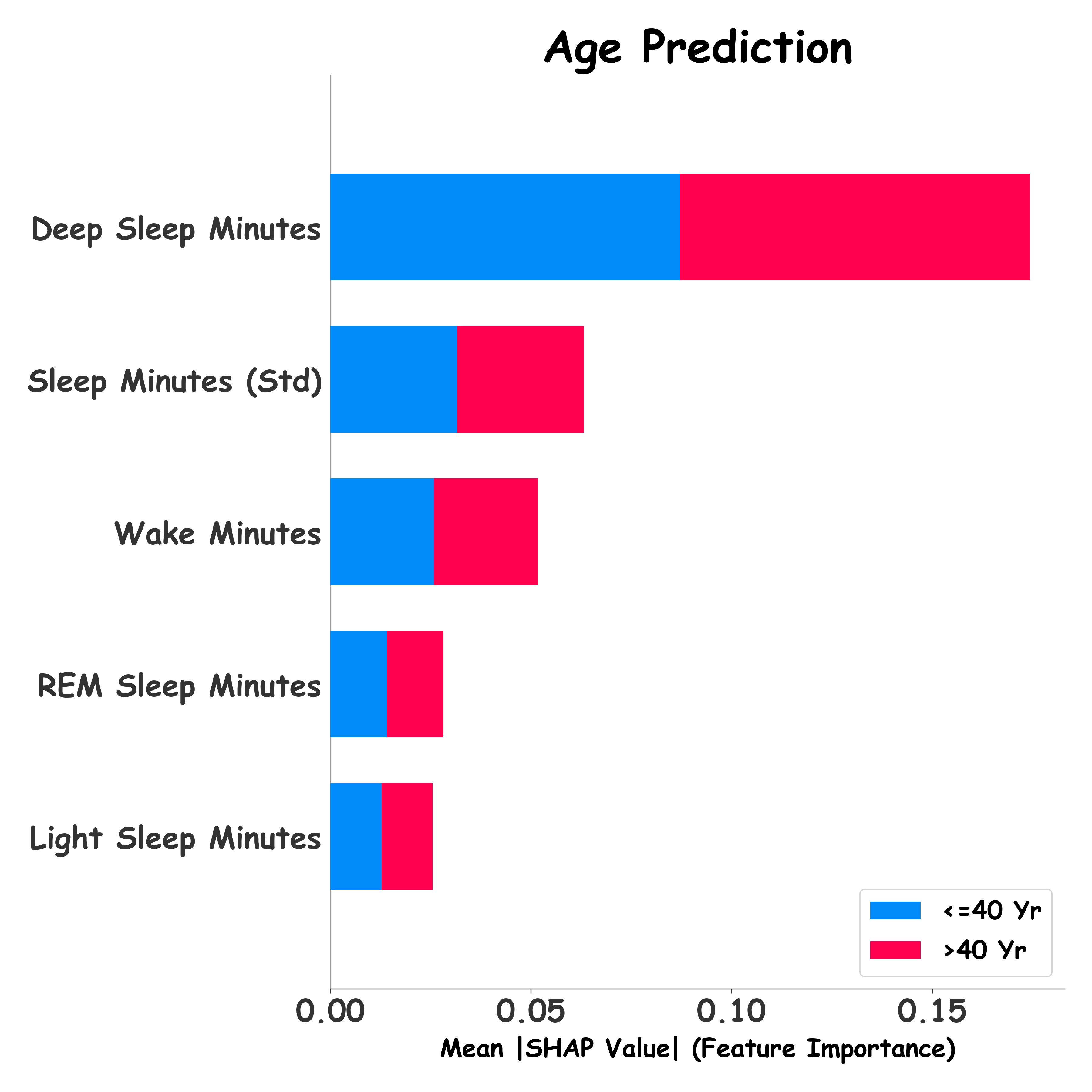}};
        \node[draw=none,fill=none] at (0.33\linewidth, 0){\includegraphics[width=0.3\linewidth]{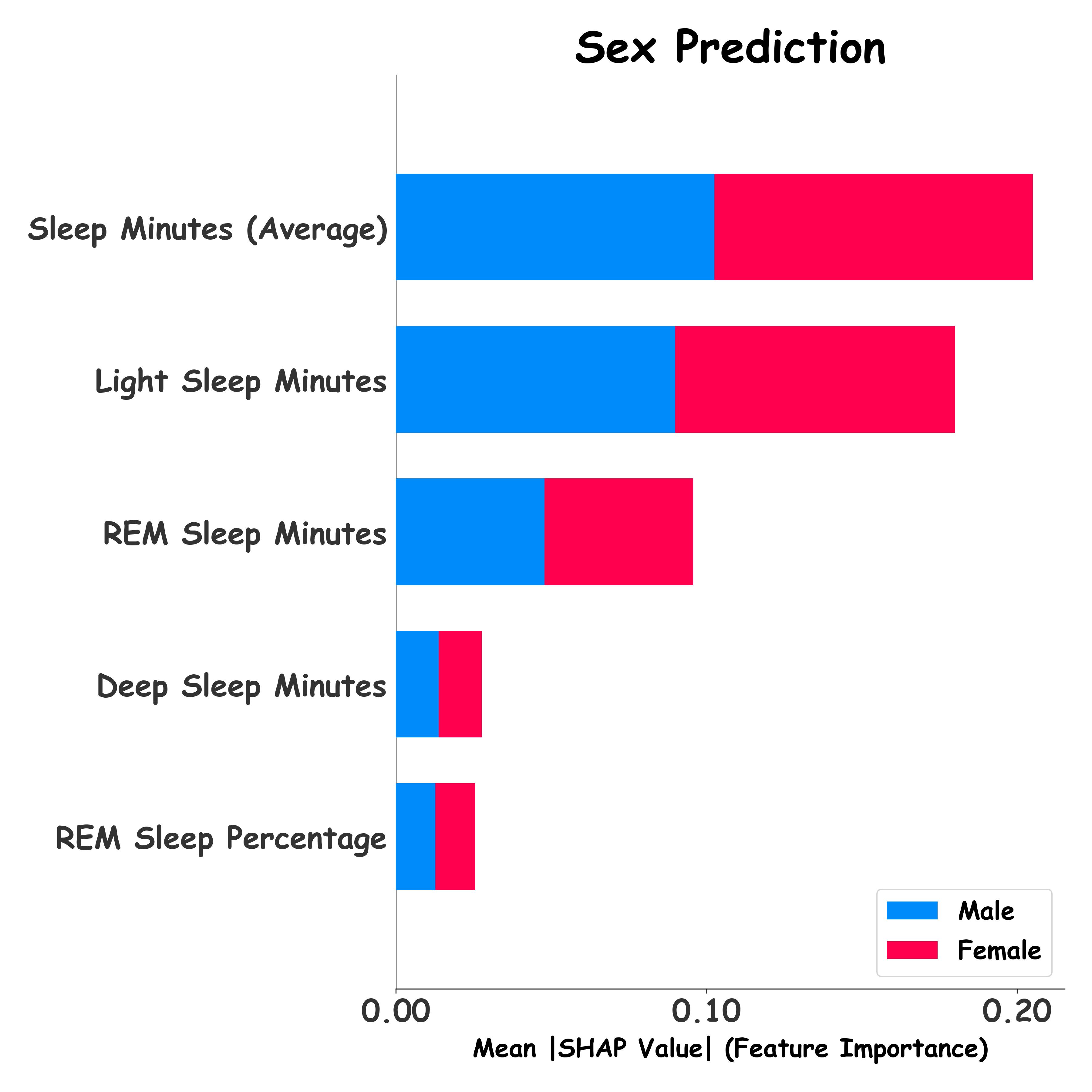}};
        \node[draw=none,fill=none] at (0.66\linewidth, 0){\includegraphics[width=0.3\linewidth]{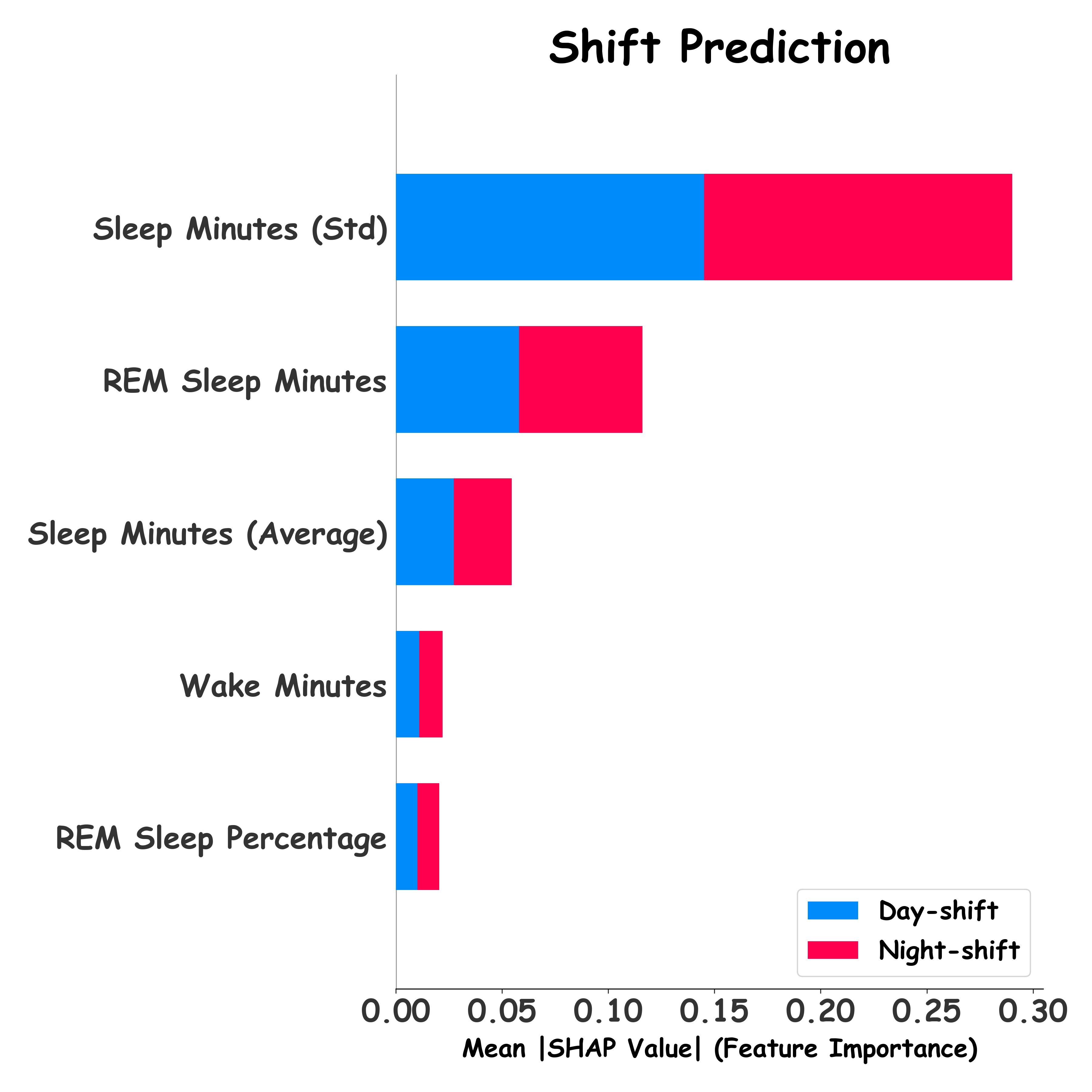}};
    \end{tikzpicture}
    
    \label{fig:feature_importance}
} \end{figure*}

\subsection{Benchmarking Demographics Prediction}

Lastly, we extended our machine learning experiments to predict demographics using the same framework used for self-reported PSQI prediction. In contrast to the PSQI results, a simple Random Forest classifier achieved competitive performance in demographic prediction tasks. In particular, sleep features derived from Fitbit data yielded ROC-AUC scores of 0.725 and 0.754 for predicting age and primary work shift, respectively. While deep time-series models slightly outperformed the Random Forest in predicting sex and shift (based on ROC-AUC), they substantially underperformed in age prediction. An ensemble of SleepNet and TimesNet slightly improved performance compared to individual models.

To better understand the predictions, we present the top 5 most important features in our Random Forest model in Figure~\ref{fig:feature_importance} using the tree explainer toolkit from Shap \cite{lundberg2020local}. The feature importance analysis revealed that deep sleep duration and total sleep duration were the most important features for predicting age and sex, respectively. 
Specifically, the prominence of deep sleep duration in predicting age aligns with prior research showing that deep sleep tends to decline with increased age \cite{li2022sleep}.
Moreover, the standard deviation of total sleep time within the participant emerged as the most important feature for predicting primary work shifts. This implicates the variability in sleep schedules often associated with shift work. Workers with non-standard shifts typically experience greater irregularity in sleep duration, making this measure a logical predictor for the primary working shift.

\section{Discussion}
\label{sec:discussion}

\subsection{Dataset Innovation}
\label{sec:innovation}

As we presented in Section~\ref{sec:related_work}, our TILES-2018 Sleep Benchmark is one of the largest wearable sleep datasets that was collected in longitudinal studies in naturalistic settings. We would highlight several unique values that the TILES-2018 Sleep Benchmark brings to the broader biomedical and healthcare research domains.

\noindent \textbf{Promoting Hospital Worker Health}. One unique value of our dataset is its focus on understanding sleep behaviors among healthcare providers. This distinguishes our dataset from other efforts, such as the All of Us program \cite{all2019all}, which seeks to enroll a large population across the United States to collect wearable data for promoting general public health. Specifically, nurses, physicians, and other hospital staff are essential to the functioning of modern healthcare systems. However, they often encounter higher levels of occupational stress, long and irregular working hours, and physically demanding tasks compared to other occupations. These factors increase their risk for disrupted sleep patterns, circadian misalignment, and chronic sleep disorders~\cite{stimpfel2020nurses}. By capturing longitudinal, real-world sleep data from this high-risk population, our dataset offers quantifiable measures that can inform workplace interventions, support evidence-based policy decisions, optimize shift scheduling, and promote the long-term well-being of hospital workers. Not only does this dataset promote the health of hospital workers, but it also has a broader impact on advancing the quality of modern medical care.

\noindent \textbf{Comprehensive Measures of Sleep Information}. Notably, existing literature on sleep quality among hospital staff primarily relies on survey-based approaches~\cite{stimpfel2020nurses, gomez2016nurses, huang2021poor}, which often lack objective measures of sleep timing and physiological patterns. In contrast, our current work, along with our previous effort in TILES-2018, provides detailed comparisons of sleep behavior across staff working different primary work shifts (day vs. night) and in various primary work units (e.g., ICU, lab, office). Objective measures of sleep can also provide details into how individuals working irregular hours, such as night shift, adopt different coping strategies to manage sleep disruptions. In addition to objective measures, we also collect self-reported PSQI surveys to assess sleep quality, which aligns with existing studies~\cite{stimpfel2020nurses, gomez2016nurses, huang2021poor} in evaluating sleep quality among hospital staff.

\noindent \textbf{Detailed Demographic Information Coverage}. Although our dataset focuses only on hospital staff, it includes detailed demographic information such as sex, age, primary work shift, and primary work unit. This information enables subgroup analyses to study how occupational and demographic factors impact sleep behavior. In addition, the collection of self-reported PSQI surveys provides a valuable subjective measure of sleep quality, supporting joint analyses of perceived and physiological sleep patterns across these demographic groups. Moreover, the detailed demographic information can be used to conduct research studies on how reliably the machine learning models perform across different demographics.

\noindent \textbf{Unique Sleep Benchmark Dataset}. Few datasets in the existing literature have been explicitly designed to serve as standardized benchmarks for modeling wearable sleep data. In comparison to the wearable-based datasets listed in Table~\ref{tab:related_work}, the TILES-2018 Sleep Benchmark dataset features its relatively large participant size and an extended study duration of sleep recordings collected in real-world settings. Combined with the previously released TILES-2018 dataset, this resource offers over 12,000 high-quality sleep episodes, making it a unique source for training and testing deep learning models on real-world physiological data. Owing to its scale, richness, and contextual metadata, we would highlight that the dataset has the potential to support community-wide challenges aimed at benchmarking machine learning methods in biomedical and sleep informatics research.

\subsection{Benchmark Performance}
\label{sec:benchmark}

In this paper, we uniquely introduce a suite of benchmark tasks, including predicting wearable-derived sleep stage labels, self-reported sleep quality, and participant demographics based on physiological sleep features and time-series sleep data. Our benchmark results reveal several important observations. First, sleep stage classification is more accurate among younger participants, likely due to the more distinct and consistent physiological signatures of sleep stages in this age group, such as clearer REM and deep sleep patterns~\cite{li2022sleep}. Second, demographic attributes such as age, sex, and primary shift schedule can be effectively predicted from sleep-related features, suggesting the presence of structured variation in sleep physiology across groups.

On the time-series modeling front, we benchmark two popular deep learning models in the field, including SleepNet (LSTMs) and the more state-of-the-art model architecture TimesNet. Our results show that TimesNet outperforms SleepNet in predicting self-reported sleep quality and participant demographics. However, TimesNet does not outperform LSTMs in sleep stage classification, potentially due to the sequential nature of sleep stage transitions where recurrent architectures like LSTMs are more well-suited. Moreover, ensembling predictions from both models further improves performance across all tasks. In contrast, LLM-based models such as BioMistral demonstrate limited zero-shot performance on prediction tasks like self-reported sleep quality. This suggests that current LLMs, while powerful in language reasoning, are not yet optimized for interpreting physiological time-series data. Surprisingly, traditional Random Forest models using knowledge-driven sleep features (e.g., total sleep time, REM duration) deliver strong performance in predicting demographics, including age, sex, and shift type. This observation implied the importance of simple but interpretable and domain-driven features in current benchmark tasks. One possible explanation is that these knowledge-driven features effectively capture physiological patterns known to different demographic groups. For example, reduced deep sleep is associated with aging~\cite{li2022sleep}, while higher variability in sleep duration is associated with night-shift work~\cite{wright2013shift}.

\subsection{Limitations and Challenges}
\label{sec:limitation}

While the TILES-2018 Sleep Benchmark dataset is one of the largest-scale, multi-week sleep datasets collected in naturalistic settings, this dataset still encounters several limitations. It is critical to highlight these challenges across data collection, data analytics, and modeling approaches, in the context of practical applications involving wearable sleep data. 

\noindent \textbf{Device Accuracy} First, one primary challenge for our dataset lies in the accuracy of the wearable device used for data collection. While the Fitbit Charge 2 was widely adopted for consumer sleep tracking, this device does not reach the same precision as clinical-grade polysomnography (PSG). For example, previous studies, such as Stucky et al.~\cite{stucky2021validation}, have shown that the Fitbit Charge 2 provides reasonably accurate estimates of heart rate and sleep timing (e.g., sleep onset and wake-up times), they find notable inaccuracies in sleep stage classification. In particular, the device tends to overestimate REM sleep duration, which introduces errors into post-analysis and downstream sleep modeling tasks. However, we would highlight that since most of our sleep behavioral analysis (e.g., comparisons between day-shift and night-shift hospital workers) is conducted between groups using the same device under similar settings, the systematic errors introduced by the device are likely to be consistent across participants. This consistency could potentially mitigate the impact of measurement error, making our presented comparisons of sleep behavior still reliable and accurate.

\noindent \textbf{Self-reported Sleep Quality Acquisition} A second limitation of our dataset collection is the lack of sleep quality labels at the individual sleep episode level. While our study collected self-reported sleep quality using the PSQI at the beginning of the study, we did not administer daily assessments, such as Ecological Momentary Assessments (EMAs), to acquire nightly perceived sleep quality as presented in \cite{ito2023effect}. The absence of this self-reported individual sleep assessment limits our ability to benchmark models for predicting sleep quality at a finer resolution.

\noindent \textbf{Handling Noisy or Missing Data} Another major challenge in our benchmark experiments is the lack of exploration into methods for handling noisy or missing data in wearable recordings \cite{van2024mitigating}. It is known that wearable devices are inherently prone to data quality issues (such as data noise or data loss) due to motion artifacts, battery depletion, and participant compliance. Although we filtered for high-quality recordings with adequate heart rate coverage in most benchmark experiments, we did not experiment with techniques to mitigate noisy or missing data. In future work, we plan to explore data imputation, data denoising, and masked time-series learning to improve modeling robustness under real-world conditions.

\section{Conclusion}
\label{sec:conclusion}

In this paper, we present a longitudinal wearable sleep dataset of hospital workers, not published before, referred to as \textit{TILES-2018 Sleep Benchmark}. This dataset includes over $5,000$ unique sleep samples from $139$ hospital employees, in conjunction with demographic data and self-reported sleep quality assessments. We conduct several distinct analyses to demonstrate the potential of the dataset, including comparisons of sleep patterns, sleep stage classification, prediction of self-reported sleep quality, and demographic classification. Our results highlight the unique opportunities that \textit{TILES-2018 Sleep Benchmark} offers to the research community for advancing research on real-world sleep behaviors. Moreover, our dataset provides a valuable resource for benchmarking computational models in sleep behavior understanding and analytics.

\section{Acknowledgement}
The research is based upon work supported by the Office of the Director of National Intelligence(ODNI), Intelligence Advanced Research Projects Activity(IARPA), via IARPA Contract No $2017$ - $17042800005$. The views and conclusions contained herein are those of the authors and should not be interpreted as necessarily representing official policies or endorsements, either expressed or implied, of ODNI, IARPA, or the U.S. Government. The U.S. Government is authorized to reproduce and distribute reprints for Governmental purposes notwithstanding any copyright annotation thereon.

\bibliographystyle{IEEEbib}
\bibliography{strings,refs}

\end{document}